\documentclass[spaces,hyphens]{llncs}
\usepackage[a-1b]{pdfx} 
\usepackage{graphicx}
\usepackage{rotating}
\usepackage{booktabs}

\begin{document}

\title{The Turing Test for Graph Drawing Algorithms}
%
%
\author{Helen C. Purchase\inst{1}\and
  Daniel Archambault\inst{2}\and
  Stephen Kobourov\inst{3}\and
  Martin~N\"ollenburg\inst{4}\and
  Sergey Pupyrev\inst{5} \and
  Hsiang-Yun Wu\inst{4}
}
\authorrunning{H. C. Purchase et al.}
%
\institute{University of Glasgow, Glasgow, UK. \email{Helen.Purchase@glasgow.ac.uk}
\and Swansea University, Swansea, UK. \email{d.w.archambault@swansea.ac.uk}
\and University of Arizona, Tucson, USA. \email{kobourov@cs.arizona.edu}
\and TU Wien, Vienna, Austria.\\ \email{noellenburg@ac.tuwien.ac.at, hsiang.yun.wu@acm.org}
\and Facebook, Menlo Park, USA. \email{spupyrev@gmail.com}
}
%
\maketitle              
\begin{abstract}
Do algorithms for drawing graphs pass the Turing Test?  That is, are their outputs indistinguishable from graphs drawn by humans? We address this question through a human-centred experiment, focusing on `small' graphs, of a size for which it would be reasonable for someone to choose to draw the graph manually. Overall, we find that hand-drawn layouts can be distinguished from those generated by graph drawing algorithms, although this is not always the case for graphs drawn by force-directed or multi-dimensional scaling algorithms, making these good candidates for Turing Test success. We show that, in general, hand-drawn graphs are judged to be of higher quality than automatically generated ones, although this result varies with graph size and algorithm.
\end{abstract}

\section {Introduction}
It is common practice to use node-link diagrams when presenting graphs to an audience (e.g.,~online, in an article, to support a verbal presentation, or for educational purposes), rather than the alternatives of adjacency matrices or edge lists. Automatic graph layout algorithms replace the need for a human to draw graphs; it is important to determine how well these algorithms fulfil the task of replacing this human activity,

Such algorithms are essential for creating drawings of large graphs; it is less clear that this is the case for drawing smaller graphs. In our experience as graph drawing researchers, it is often preferable to draw a small graph ourselves, how we wish to depict it, than be beholden to the layout criteria of automatic algorithms. 

The question therefore arises: are automatic graph layout algorithms any use for small graphs? Indeed, for small graphs, is it even possible to tell the difference? If automatic graph layout algorithms were doing their job properly for small graphs, then they should produce drawings not dissimilar to those we would choose to create by hand.

Distinguishing human and algorithmic graph drawings can be considered a `Turing Test'; as in Turing's 1950 `Imitation Game'~\cite{machinery1950computing}, if someone cannot tell the difference between machine output and human output more than half the time, the machine passes the Turing Test. Thus, if someone cannot tell the difference between an algorithmically-drawn graph and a hand-drawn graph more than half the time, the algorithm passes the Turing Test: it is doing as good a job as human graph drawers. Of course, algorithms are useful for non-experts and for large graphs that cannot be drawn by humans effectively, but in the context of experts presenting a small graph, can their creations be distinguished from products from layout algorithms? Turing Tests have never yet been performed on graph layout algorithms. 

This paper presents the results of an experiment where participants were asked to distinguish between small hand-drawn graphs and those created by four common graph layout algorithms. Using different algorithms and graphs of different size allows us to investigate under what conditions an algorithm might pass the Turing Test. Our Turing Test results led us to also ask, in common with the \textit{Non-photorealistic rendering Turing Test} observational study~\cite{incsj-nrcos-06}, which of the two methods of graph drawing (by hand, or by algorithm) produce better drawings. We find that distinguishing hand-drawn layouts from automatically generated ones depends on the type of the layout algorithm, and that subjectively determined quality depends on graph size and the type of the algorithm.

\section{Related Work}

\subsection{Automatic Graph Layout algorithms}\label{sec:related_algorithms}

We focus on four popular families of layout algorithms \cite{Battista:1998:GDA,Gibson:2012:IV}: force-directed, stress-minimisation, circular and orthogonal.

Most general-purpose graph layout algorithms use a force-directed (FD)~\cite{Eades+1984a,Fruchterman_Reingold_1991} or stress model~\cite{brandes2007,koren2002ace}. FD works well for small graphs, but does not scale for large networks. Techniques to improve scalability often involve multilevel approaches, where a sequence of progressively coarser graphs is extracted from the graph, followed by a sequence of progressively finer layouts, ending with a layout for the entire graph~\cite{10Bartel,gk-grip-00,hh-msadg-99,harelkoren02,hu2006}.

Stress minimisation, introduced in the general context of multi-dimensional scaling (MDS)~\cite{MDS} is also frequently used to draw graphs~\cite{89Kamada,seery1980designing}.
Simple stress functions can be optimised by exploiting fast algebraic operations such as majorisation. Modifications to the stress model include the strain model (classical scaling)~\cite{torgerson1952}, PivotMDS~\cite{brandes2007}, COAST~\cite{gansner2013coast}, and MaxEnt~\cite{gansner2013maxent}.

Circular layout algorithms~\cite{st-cda-13} place nodes evenly around a circle with edges drawn as straight lines. Layout quality (in particular the number of crossings) is influenced by the order of the nodes on the circle. Crossing minimisation in circular layouts is NP-hard~\cite{mnkf-cmleg-90}, and various heuristics attempt to find good vertex orderings~\cite{bb-crcl-04,gk-icl-06,kmn-eebda-17}. 

The orthogonal drawing style~\cite{efk-ogd-01} is popular in applications requiring a clean and schematic appearance (e.g., in software engineering or database schema). Edges are drawn as polylines of horizontal and vertical segments only. Orthogonal layouts have been investigated for planar graphs of maximum degree four~\cite{t-eggmn-87}, non-planar graphs~\cite{bk-bhogd-98} and graphs with nodes of higher degree~\cite{bmt-tmuaogd-00,fk-aabnod-97}.

We seek to understand if drawings produced by these types of algorithms can be distinguished from human-generated diagrams for small networks. We do this by asking experimental participants to identify the hand-drawn layout when it is paired with an algorithmically-created one.

\subsection{Studies of Human-Created Graph Layouts}\label{sec:related_studies}

Early user studies~\cite{pcj-vgda-96,p-wageh-98} confirmed that many of the aesthetic criteria incorporated in layout algorithms (e.g., uniform edge length, crossing minimisation) correlate with user performance in tasks such as path finding. Van Ham and Rogowitz~\cite{hr-pougl-08} investigated how humans modified given small graph layouts so as to represent the structure of these graphs. They found that force-directed layouts were already good representations of human vertex distribution and cluster separation.  Dwyer et al.~\cite{dlfqir-cuagl-09} focused on the suitability of graph drawings for four particular tasks (identifying cliques, cut nodes, long paths and nodes of low degree), and found that the force-based automatic layout received the highest preference ratings, but the best manual drawings could compete with these layouts. Circular and orthogonal layouts were considerably less effective.  Purchase et al.~\cite{ppp-gdacua-12} presented graph data to  participants as adjacency lists and asked them to create drawings by sketching; their findings include that the participants preferred planar layouts with straight-line edges (except for some non-straight edges in the outer face), nodes aligned with an (invisible) grid, and somewhat similar edge lengths.  Kieffer et al.~\cite{kdmw-hhonl-16} focused on orthogonal graph layouts, asking participants to draw a few small graphs (13 or fewer nodes) orthogonally by hand. The human drawings were compared to orthogonal layouts generated by yEd~\cite{yed} and the best human layouts were consistently ranked better than automatic ones. They then developed an algorithm for creating human-like orthogonal drawings.


This paper builds on this prior work  by considering drawings of small to medium-sized graphs (up to 108 nodes) and an example from each of four families of standard graph layout algorithms. We address the question of whether people can distinguish between algorithmic and human created drawings, and if so, is this the case for all layout algorithms?   

\begin{table*}[tb]
\caption{
Characteristics of the experimental graphs. The \emph{size} column indicates how the graphs were divided into sub-sets (small, medium, large) for the purposes of the experiment; (rw): real-world graphs; (ab): abstract graphs.
}
\scriptsize
\begin{tabular}{llllp{11.5mm}p{12.5mm}lllp{40mm}} \toprule
graph & nodes & edges	& density	& mean shortest path & clustering coefficient & diam. & planar & size & reference \\ \midrule
$G_1$(rw)		& 108	& 156	& 0.03	& 5.03	& 0.11	& 11 & N	& L 	& {Causes of obesity~\cite{g1Paper}}\\ 
$G_2$(rw)		& 22	& 164	& 0.71	& 1.30	& 0.78	& 2	 & N 	& S 	& {Causes of social problems in Alberta, Canada~\cite{g2}}\\ 
$G_3$(rw)		& 85	& 104	& 0.03	& 6.05	& 0.04	& 13 & Y 	& L 	& {Cross posting users on a newsgroup (final timeslice)~\cite{08FrishmanOnlineTVCG}}\\ 
$G_4$(rw)		& 34	& 77	& 0.14	& 2.45	& 0.48	& 5  & N 	& M 	& {Social network~\cite{karateClub}}\\ 
$G_5$(ab)		& 20	& 30	& 0.16	& 2.63	& 0.00	& 5  & Y 	& S 	& {Fullerene graph with 20 nodes~\cite{g5}}\\ 
$G_6$(ab)		& 24	& 38	& 0.14	& 3.41	& 0.64	& 6  & N 	& S 	& {A block graph (chordal, every biconnected component is a clique)~\cite{g6}}\\ 
$G_7$(ab)		& 42	& 113	& 0.13	& 2.55	& 0.48	& 5  & Y 	& M 	& {A maximal planar graph~\cite{g7}}\\ 
$G_8$(ab)		& 37	& 71	& 0.11	& 2.76	& 0.70	& 5  & Y 	& M	& {A planar 2-tree~\cite{g9}}\\ 
  $G_9$(ab)	& 18	& 27	& 0.18	& 2.41	& 0.00	& 4  & N 	& S		& {Pappus graph (bipartite, 3-regular)~\cite{g10}}\\ \midrule
\textit{mean}	& 43.3	& 86.7	& 0.18	& 3.18	& 0.36	& 6.2 \\ 
\textit{median}	& 34		& 77		& 0.14	& 2.63	& 0.48	& 5\\ \bottomrule
\end {tabular}
\label{table:char}
\end{table*}

\section{Experiment}\label{sec:method}
%

\subsection{Stimuli}

\subsubsection{The Graphs.}

Our experiment compares unconstrained hand-drawn graphs with the same graphs laid out using different algorithmic approaches. We considered 24 graphs, from which we selected 9, based on the following criteria:

\begin{itemize}
	\item	A balanced split between real-world graphs and abstract graphs, the abstract graphs being ones of graph-theoretic interest;
	\item	A balanced split between planar and non-planar graphs;
	\item	A range in the number of nodes between $15$ and $108$;
	\item	A range in the number of edges (for our graphs, between 27 and 164);
	\item	Connected and undirected graphs only: directionality was removed from the real-world graphs as necessary.
\end{itemize}

The diversity of our graphs is demonstrated by the range of values for other graph characteristics (diameter, density, average shortest path length, clustering coefficient) that they exhibit (Table~\ref{table:char}).

\subsubsection{The Algorithms.}
We included examples of major families of graph drawing algorithms (Table~\ref{tableGAUsed}: force-directed, stress-based, circular, orthogonal), as implemented in yEd~\cite{yed} and GraphViz~\cite{graphviz}. HOLA~\cite{kdmw-hhonl-16} was considered, but its orthogonal design was deliberately based on human preferences (unlike the other algorithms), and so its inclusion would introduce a bias that could distort human judgements. We considered structure-specific algorithms (e.g., algorithms designed for planar graphs or trees), but for generality used generic algorithms that could handle all nine graphs, leaving specific algorithms for future work. 


\begin{table}[tb]
\caption{The four graph layout algorithms used.}
\footnotesize
\label{table:alg}
\centering{
\footnotesize
\begin{tabular}{llll}
	\toprule
	algorithm ID & algorithm type & original name & parameters \\
	\midrule
$A_{\mathrm{FD}}$	& force-directed			& Organic~\cite{yed}	& default 	 		\\ 
$A_{\mathrm{MDS}}$	& stress-based	& MDS~\cite{graphviz}	& default	\\ 
$A_C$	& circular					& Circular~\cite{yed}		& default	 		\\ 
$A_O$	& orthogonal				& Orthogonal~\cite{yed}	& classic, default	\\ \bottomrule
\end{tabular}
}
\label{tableGAUsed}
\end{table}

\subsubsection{The Hand-Created Drawings.}

The process of creating hand-drawn graphs mimicked the context of a graph drawing researcher deciding whether to manually draw a small graph, or to use a well-established graph layout algorithm. Thus, the graphs were drawn in the knowledge they would compete against drawings created by algorithms, making the Turing test as hard as possible. This process was therefore a mini-experiment, with four of the authors (all with graph drawing expertise, called the `drawers', $D_1$-$D_4$) as participants, the context of the study being clear to them. While the drawers might have recognised some of the graphs they were asked to draw, this scenario is comparable to a real-world situation where graph drawing researchers might  know the nature of the graph to be drawn.

The first author asked the drawers to lay out the graphs using yEd~\cite{yed}, starting from a random layout (the yEd `Random' tool). There were no other instructions: it was not specified, for example, that edges needed to be straight lines rather than splines or multiple segments, nor that nodes should not overlap, nor edges cross over nodes. To improve ecological validity, all drawers were told that they could use yEd tools to support their drawing process if they wished (as likely to happen in practice). However, somewhat surprisingly, they all drew the graphs without any yEd tool support (automatic layout or otherwise). The drawers suggested doing the exercise again on a `manually-adjusted' basis; that is, using the output from a yEd layout algorithm of their choice as an initial starting point.  However, once  we paired the algorithmic drawings with their manually-adjusted versions, most of them were visually almost identical.  We therefore only used the initial hand-drawn versions.  


The mini-experiment output is a set of visual stimuli comprising $9$ graphs ($G_1, ..., G_9$), each with four layout algorithms applied $G_1A_{\mathrm{FD}}$, $G_1A_{\mathrm{MDS}}$, \dots, $G_2A_{\mathrm{C}}$, \dots, $G_9A_O$) and each with four hand-drawn versions ($G_1D_1$, $G_1D_2$, \dots, $G_2D_1$, \dots, $G_9D_4$), all represented in yEd. All $72$ drawings were subject to the same automatic scaling process to ensure the same vertex size and edge thickness. After scaling, all drawings were automatically converted into jpeg images.

\subsection{Experimental Design}

\begin{figure}[t]
\centering{
 	\includegraphics[width=0.7\linewidth]{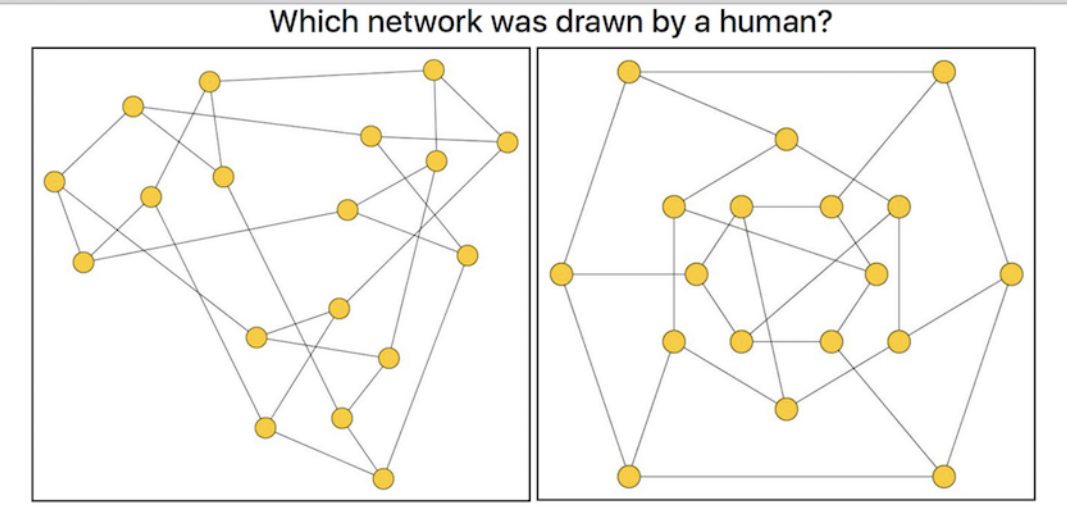} 
}
 \caption{Screen shot of the experimental system.}
 \label{fig:screen}
\end{figure}

Each experimental trial (Fig.~\ref{fig:screen}) comprises two versions of the same graph, one hand-drawn, and one created by a layout algorithm. For each graph, we firstly paired the four algorithmic versions (on the left) with the four hand-drawn versions (right) ($16$ pairs). We then flipped the algorithmic versions along the $y$ axis (reducing the possibility of participants remembering the algorithm drawings), and paired the flipped versions (right) with the four drawn versions (left) ($32$ pairs for each graph). Putting all graphs in one experiment means $288$ trials, an unreasonably long experiment. The alternative of running a separate experiment for each graph means several very small experiments, greatly increasing the number of participants needed. As a compromise, we divided our $9$ graphs into three sets, (loosely `small', `medium' and `large' (Table~\ref{table:char})), a convenience decision so as to reduce the duration of each experiment while ensuring we would be able to recruit enough participants. We thus had three sub-experiments, one `small' ($128$ trials), one `medium' ($96$ trials) and one `large' ($64$ trials).  

Using a custom-built online experimental system, participants read instructions and information about graphs (referred to as `networks') and indicated consent before proceeding. They were told it would always be the case that the two drawings presented were the same graph. Twelve practice trials used a different graph of similar size for 
familiarisation purposes.  Experimental trials were presented in random order, with no distinction between graphs. Participants took a self-timed break every $20$ trials, and demographic data was collected. 

\section{Results and Data Analysis}\label{sec:result}
The experimental link was distributed to authors' colleagues, students, family and friends. Participants were considered outliers if their mean time over all trials was unreasonably low (less than 1 second, $n=2$), or if they consistently responded one side for a large number of consecutive trials (e.g., always left, $n=1$). No participants consistently alternated left and right. We removed the data from one participant who used a very small screen ($198 \times 332$ pixels), unconvinced that the stimuli could be perceived sufficiently well. Although some participants did not complete the experiment, since the answer to each trial is a data point in its own right (i.e.,~it is  independent and its value to the experiment does not depend on answers to any other trial), we retained all data for participants who completed at least $3/4$ of the trials, inferring that those who did not do so ($n=20$) might not have taken the experiment seriously.

Data from $46$ participants was analysed; a total of 4364 independent decisions. We categorised participants as expert ($n=21$) if their self-declared knowledge of network drawings was `expert', `highly knowledgeable', or `knowledgeable', and novice ($n=22$) for `somewhat knowledgeable' or `no knowledge'.  Three participants did not provide full demographic details.

\subsection{Data Analysis Methods}

Our data was analysed in three parts:
Part $1$ investigates the extent to which `human' was chosen over `algorithm', comparing the proportion of responses with random selection. We look at overall responses, responses for each algorithm, for each graph size, for novice and expert participants, for planar and non-planar graphs, and consider the combination of graph size and algorithm. The Binomial distribution test compares observed proportion against the `random' proportion of 0.5, where each trial is independent; its calculated p-value represents the probability that the mean of the population distribution (based  on the observed samples) is equal to 0.5. A p-value $<0.05$ indicates a significant result: that is, the observed choice proportion is so much greater than 0.5 that there is a very low  probability that the hand-drawn and algorithmically drawn graphs cannot be distinguished; statistically, this means there is insufficient evidence to indicate Turing Test success. A p-value $>0.05$ is a high probability that hand-drawn and algorithmically drawn graphs cannot be distinguished: thus, Turing Test success. We apply p-value Bonferroni corrections when dividing the data sets.

Part $2$ considers response times with respect to different algorithms, sizes, expertise, and planarity, using non-parametric tests since our data is not normally distributed. Response time is considered as a proxy for the perception of difficulty of the task: participants will take longer if they find the task difficult.

Part $3$ identifies trials with extreme responses (high or low response time, or extreme proportional choice).

A choice for a hand-drawn graph is scored as $1$; a choice for an algorithmic drawings is $0$. Thus, proportions $>0.5$ indicate that the human drawing was selected more often on average. Proportions $< 0.5$ indicate that the algorithmic drawing was (incorrectly) selected with greater frequency. 

\subsection{Results}

\subsubsection{Choice of drawing. }

Our hypotheses are:
\begin{itemize}
\item	$H_0$: It is not possible to distinguish algorithmic drawings from hand-drawn ones; thus, the true proportion $=0.5$; the algorithm passes the Turing test. This hypothesis is accepted if the Binominal p-value $>0.05$.
\item	$H_1$: It is possible to distinguish algorithmic drawings from hand-drawn ones; thus, the true proportion $\ne 0.5$. \end{itemize}

Binomial test results over all 4364 data points are shown in Table~\ref{table:expA}. Accepting $H_0$ means it is not possible to distinguish between hand-drawn and algorithmic drawings: the Turing Tests succeeds. Rejecting it means that there is insufficient support for the hypothesis; we infer that telling the difference is possible. There are no proportions $<0.5$, so no cases where, on average, algorithmically-drawn graphs were incorrectly selected more often than hand-drawn ones.

\begin{table*}[t]
\caption{
Binomial test results for `Which network was drawn by a human?' Accepting $H_0$ indicates Turing Test 'pass'. Although $0.049 < 0.05$, statistical correction means the MDS p-value threshold is $0.05/4 = 0.0125$. The corrected Novice p-value threshold is $0.05/2=0.025$, a significant result.}
  \centering
  \scriptsize
\begin{tabular}{lp{16mm}p{12mm}p{14mm}p{14mm}p{14mm}} \toprule
    & \raggedright Number of samples & Mean response time (s) & Observed proportion & Binomial p-value & Result	 \\ \midrule
All trials & $4364$ & $3.14$	& $0.56$	& $p<0.001$	& reject $H_0$	\\ 
Force-Directed ($A_{\mathrm{FD}}$)	& $1094$ & $4.26$	& $0.51$	& $p=0.566$	& accept $H_0$	\\ 
MDS ($A_{\mathrm{MDS}}$)				& $1090$ & $3.32$	& $0.53$	& $p=0.049$	& reject $H_0$	\\ 
Circular ($A_C$)		& $1090$ & $2.85$	& $0.56$	& $p<0.001$	& reject $H_0$	\\ 
Orthogonal ($A_O$)		& $1090$ & $2.79$	& $0.65$	& $p<0.001$	& reject $H_0$	\\ \midrule

Small graphs ($G_2$, $G_5$, $G_6$, $G_9$) & $1656$ & $2.58$ & $0.55$ & $p<0.001$ & reject $H_0$\\ 
Medium graphs ($G_4$, $G_7$, $G_8$) & $1817$ & $3.08$ & $0.55$ & $p<0.001$ & reject $H_0$\\ 
Large graphs ($G_1$, $G_3$) & $891$ & $4.28$ & $0.62$ & $p<0.001$ & reject $H_0$\\ \midrule
Expert participants	& $1915$ & $3.99$	& $0.63$	& $p<0.001$	& reject $H_0$	\\ 
Novice participants	& $2101$ & $2.74$	& $0.53$	& $p=0.016$	& reject $H_0$	\\ 
Planar graphs		& $2069$ & $3.15$	& $0.55$	& $p<0.001$	& reject $H_0$	\\ 
Non-planar graphs	& $2295$ & $3.49$	& $0.58$	& $p<0.001$	& reject $H_0$	\\ \bottomrule
\end{tabular}
\label{table:expA}
\end{table*}

\begin{table*}
\caption {Binomial test results by graph size and algorithm; * indicates responses sufficiently close to random for Turing Test `pass'.}
  \centering
  \scriptsize
\begin{tabular}{llllllllll} \toprule
  &\multicolumn{2}{c}{Force-Directed} &\multicolumn{2}{c}{MDS} &  \multicolumn{2}{c}{Circular} &  \multicolumn{2}{c}{Orthogonal}\\
  &proportion &p-value &proportion &p-value&proportion &p-value&proportion &p-value\\\midrule
  small & 0.52* & 0.432 & 0.57* & 0.006 & 0.51* & 0.786 & 0.62 & $<0.001$\\
  medium & 0.49* & 0.851 & 0.52* & 0.542 & 0.53* & 0.205 & 0.64 & $<0.001$\\
  large & 0.52* & 0.640 & 0.49* & 0.789 & 0.73 & $<0.001$ & 0.74 & $<0.001$\\\bottomrule
\end {tabular}
\label {sizeAlgTuring}
\end {table*}

The results indicate that people can distinguish between algorithmic and hand-drawn graphs (over all graphs and algorithms), correctly choosing the hand-drawn graph 56\% of the time ($p<0.001$). This result applies equally well regardless of graph size, viewer expertise, or graph planarity: the tests all reveal significant difference between the observed proportion and 0.5. Thus, overall, the Turing test fails.

There is a difference, however, when the algorithm is taken into account: the observed proportion for Force-Directed algorithm trials was 0.51, sufficiently close to the random response proportion of 0.50 that we can accept $H_0$, and state that this algorithm passes the Turing Test. The proportion of 0.53 for MDS is very close (but not really close enough in statistical terms), and we clearly reject $H_0$ for circular and orthogonal algorithms.

The size/algorithm combination (threshold p-value = 0.05/12 = 0.0042) reveals additional results according to the size of the graph (Table~\ref{sizeAlgTuring}). As expected, the Force-Directed algorithm gives proportions close to 0.5 for all graph sizes. The MDS results suggest Turing Test success for all three sizes when analysed separately (albeit a marginal result for the smallest graphs), even though the overall MDS result reported above (at $p=0.049$) indicates rejection of the null hypothesis. The MDS result is therefore clearly on the boundary of success. There are Turing Test passes for small and medium graphs for the Circular algorithm.

\subsubsection{Response Time.}

Non-parametric tests on response time for algorithm and graph size (Table~\ref{table:expA}) reveals that MDS decisions were slower than orthogonal ones (adjusted pairwise comparison after repeated measures Freidman, $p=0.022$), decisions on large graphs were slower than on small graphs (adjusted pairwise comparison after independent measures Kruskal Wallis, $p=0.039$), and experts made slower decisions than novices (independent measures Mann-Whitney, $p=0.014$). There was no statistical difference between response times with respect to graph planarity.

\subsubsection{Extreme Examples}

\begin{figure}
\centering
\scriptsize
\centering
\begin{tabular}{lccccc} \toprule
  \multicolumn{6} {c} {\textbf {Slow Response Time (Seconds)}}\\\midrule
   & $G_4A_{\mathrm{MDS}}$/$G_4D_4$ & $G_7A_C$/$G_7D_4$ & $G_3A_{\mathrm{FD}}$/$G_3D_2$ & $G_3A_{\mathrm{FD}}$/$G_3D_1$ & $G_3A_{\mathrm{FD}}$/$G_3D_4$ \\
   & ($47.88s$) & ($48.09s$) & ($48.32s$) & ($49.49s$) & ($50.03s$) \\
  & prop $=0.40$ & prop $=0.45$ & prop $=0.46$ & prop $=0.43$ & prop $=0.61$\\
    \begin{sideways}\textbf{Algorithm} \end{sideways} & \includegraphics[width=18mm]{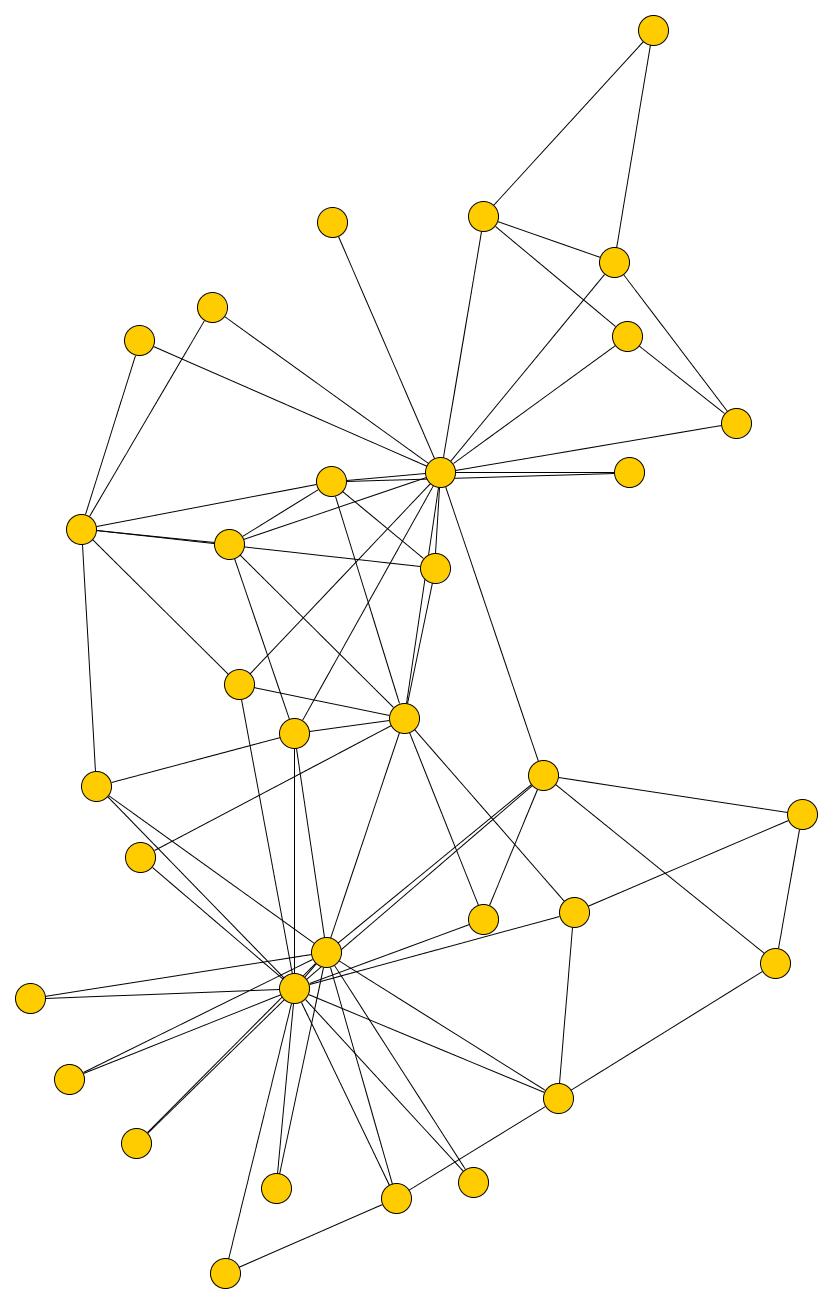} & \includegraphics[width=23mm]{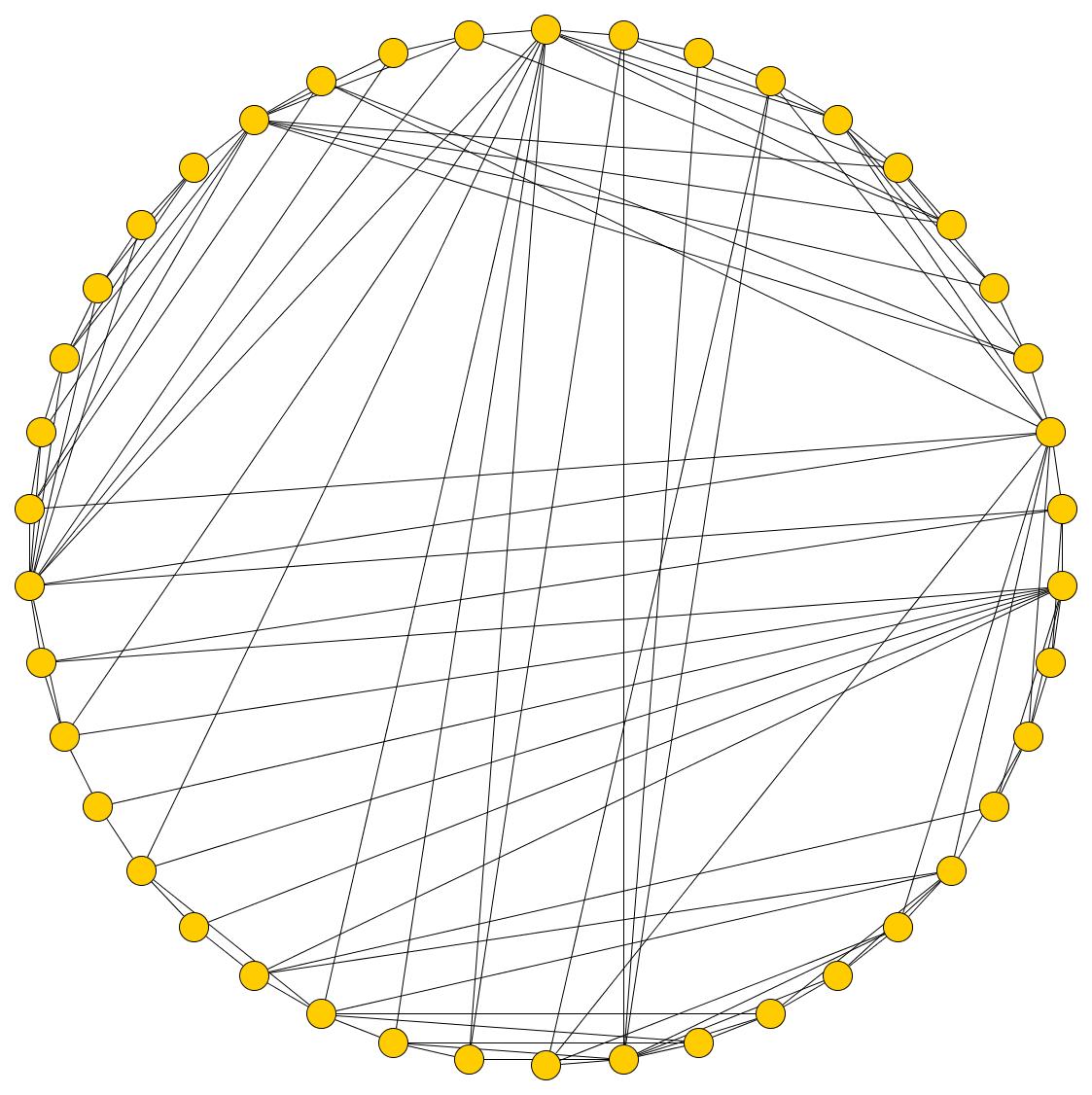} & 
    \includegraphics[width=19mm]{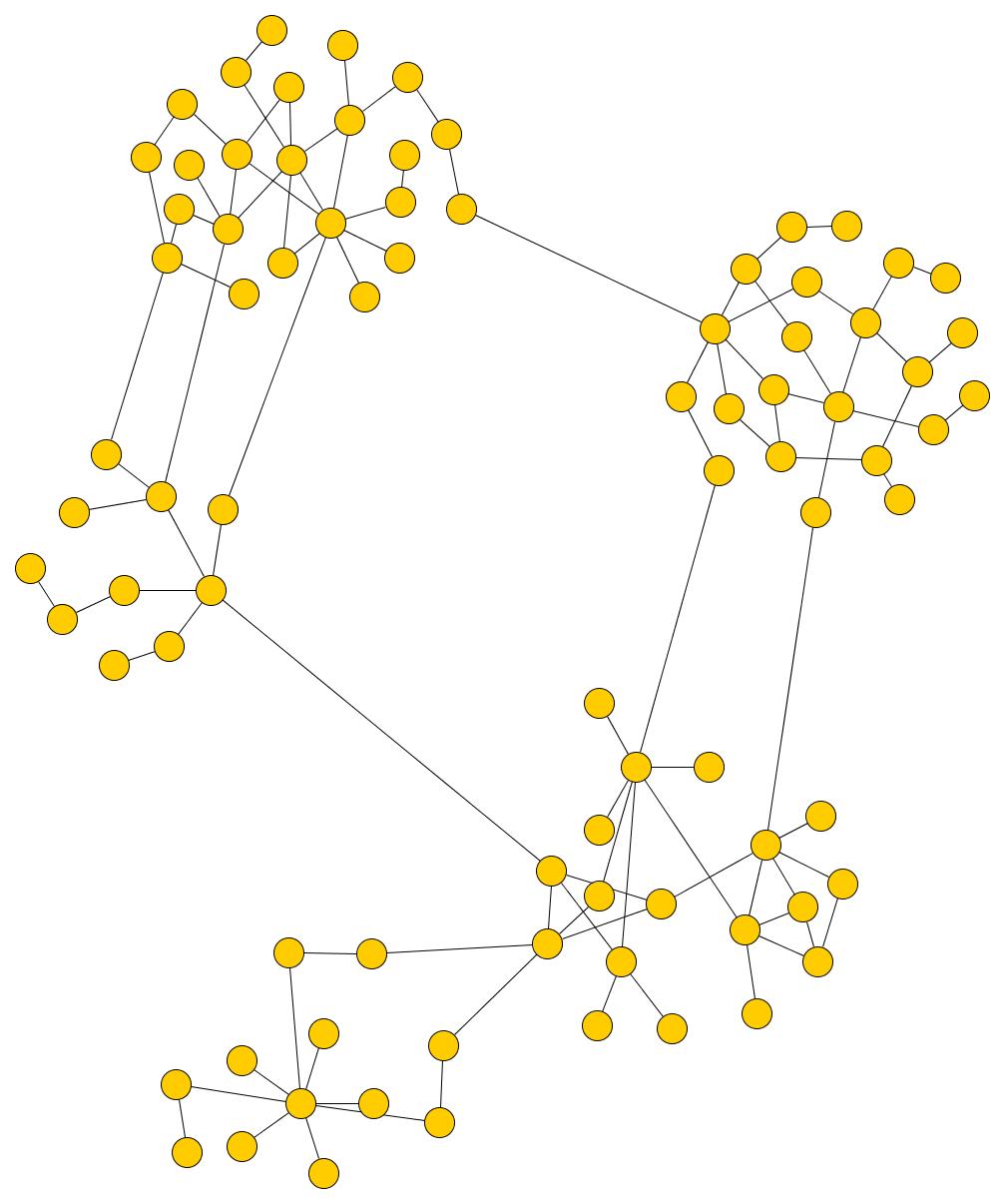} & 
    \includegraphics[width=19mm]{figures/g3a1} & \includegraphics[width=19mm]{figures/g3a1}\\
    \begin{sideways}\textbf {Human}\end{sideways} & \includegraphics[width=23mm]{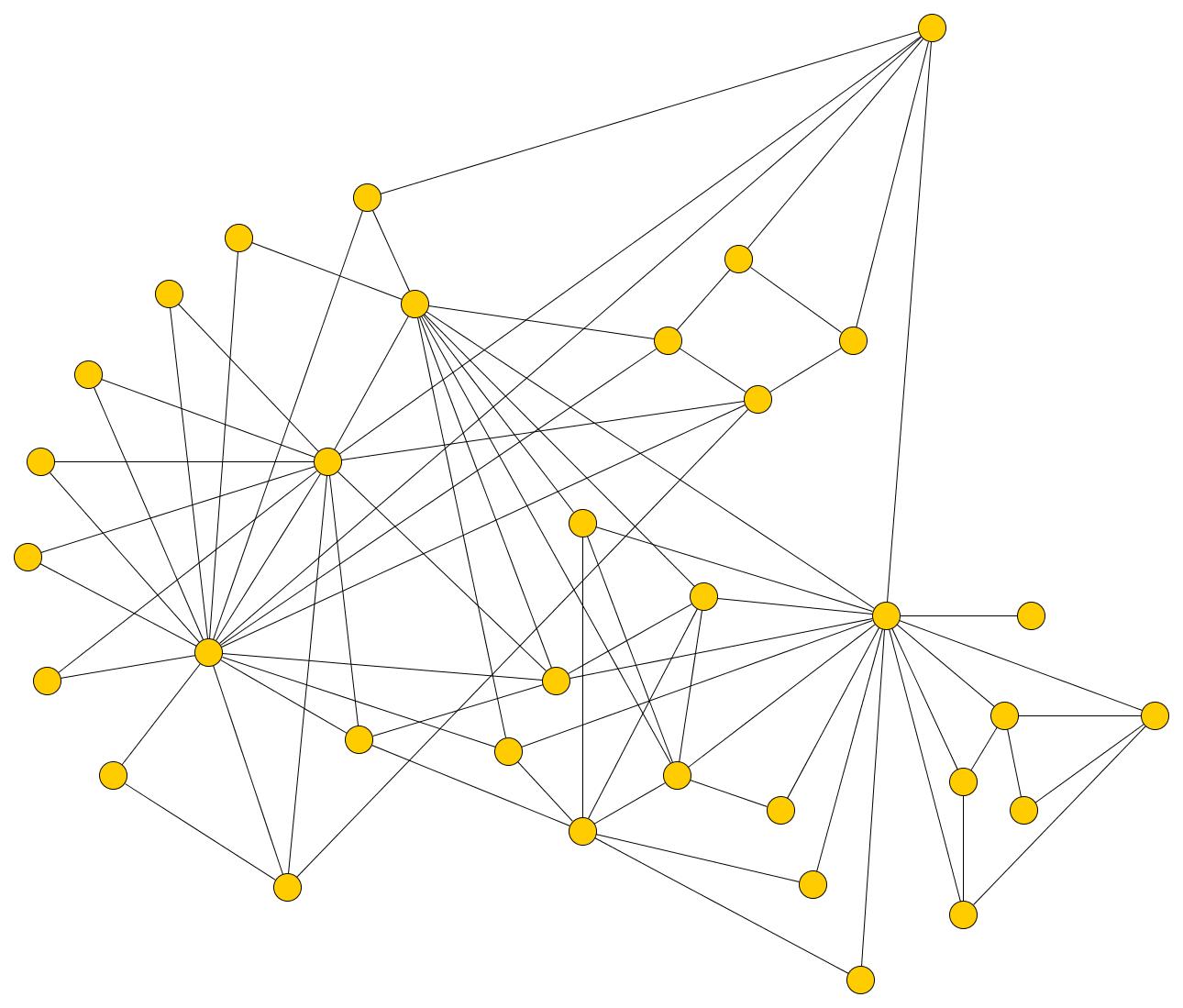} & \includegraphics[width=24mm]{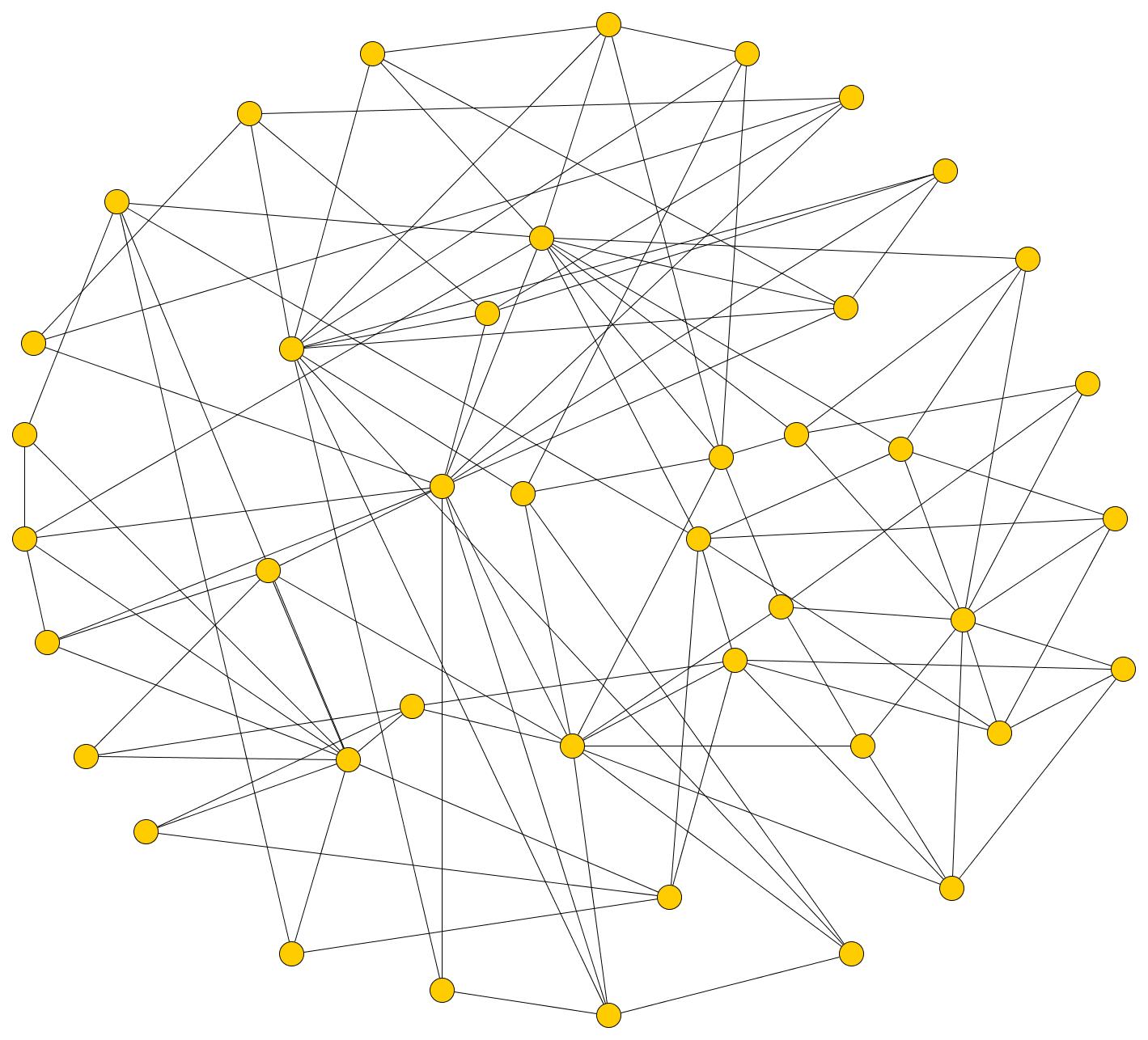} & \includegraphics[width=13mm]{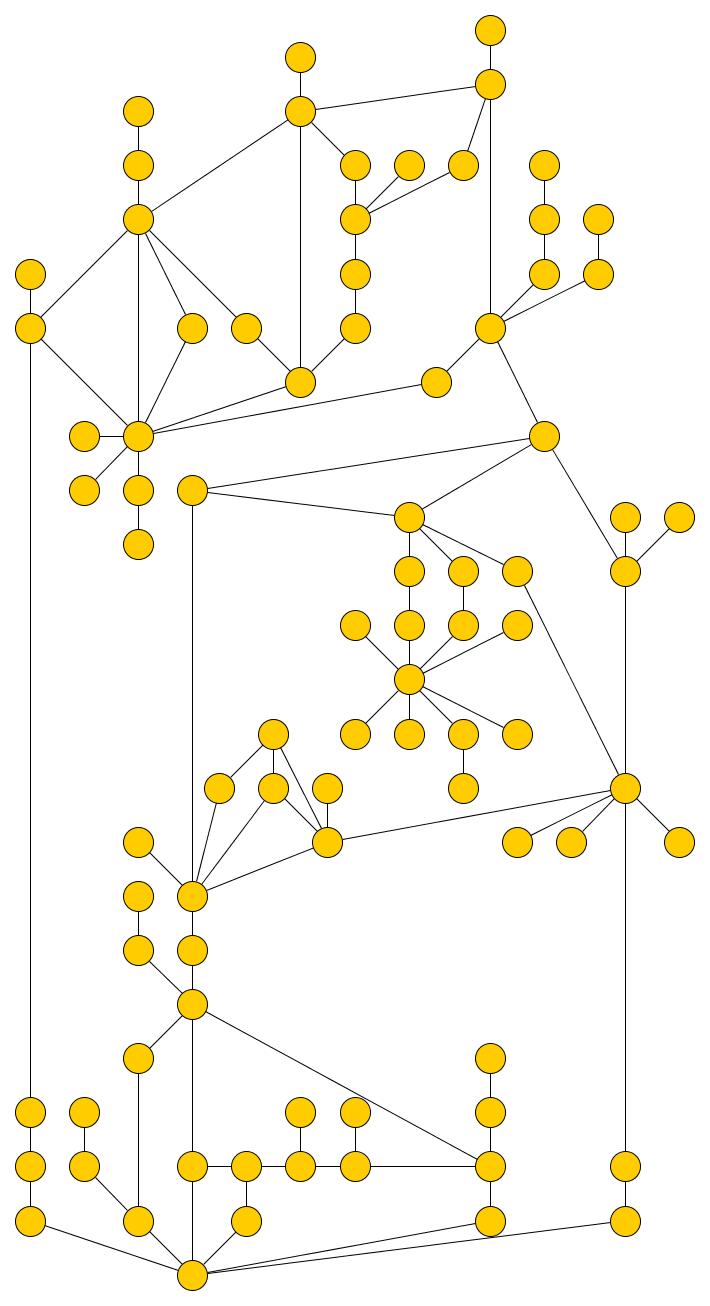} & \includegraphics[width=24mm]{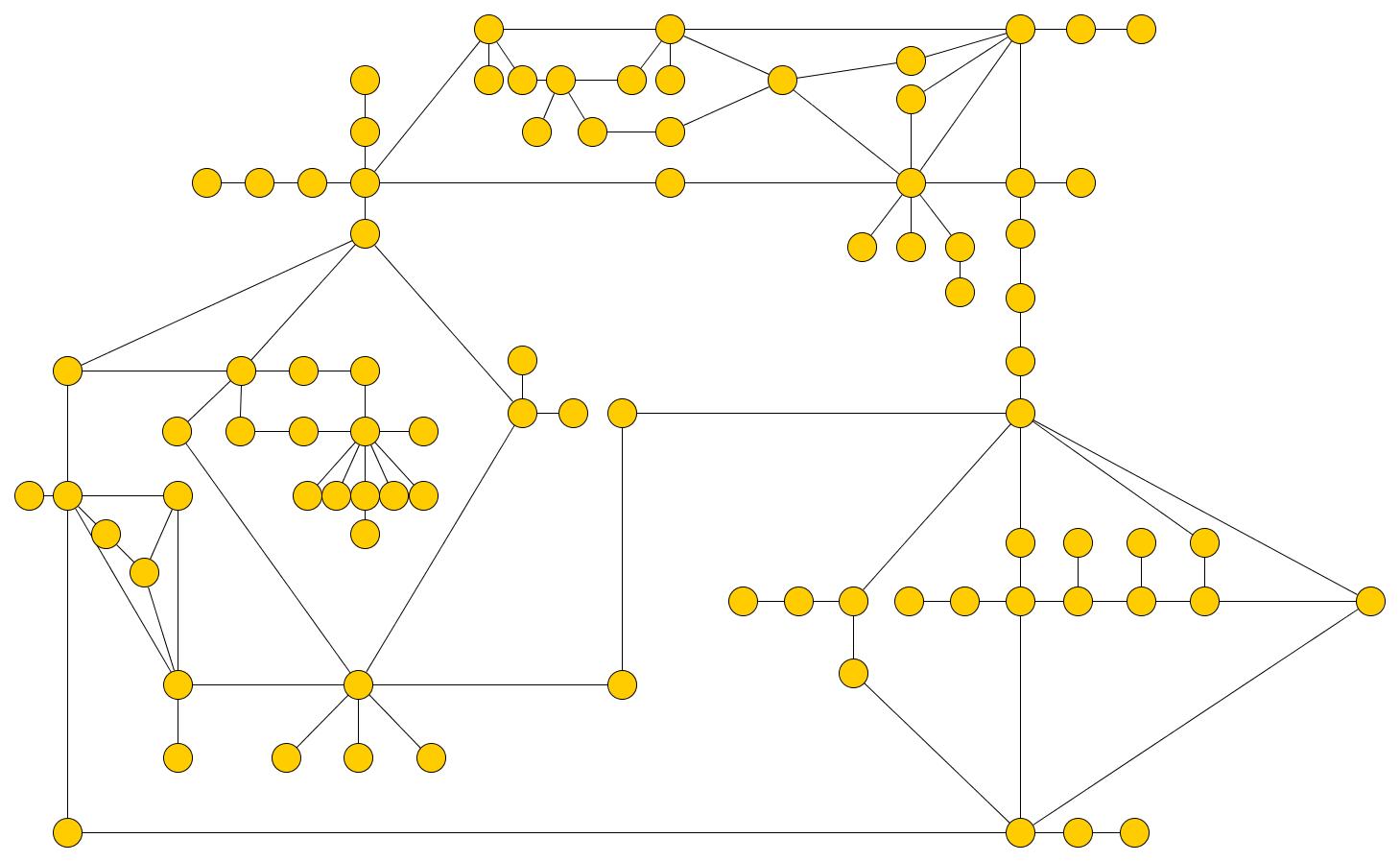} &\includegraphics[width=24mm]{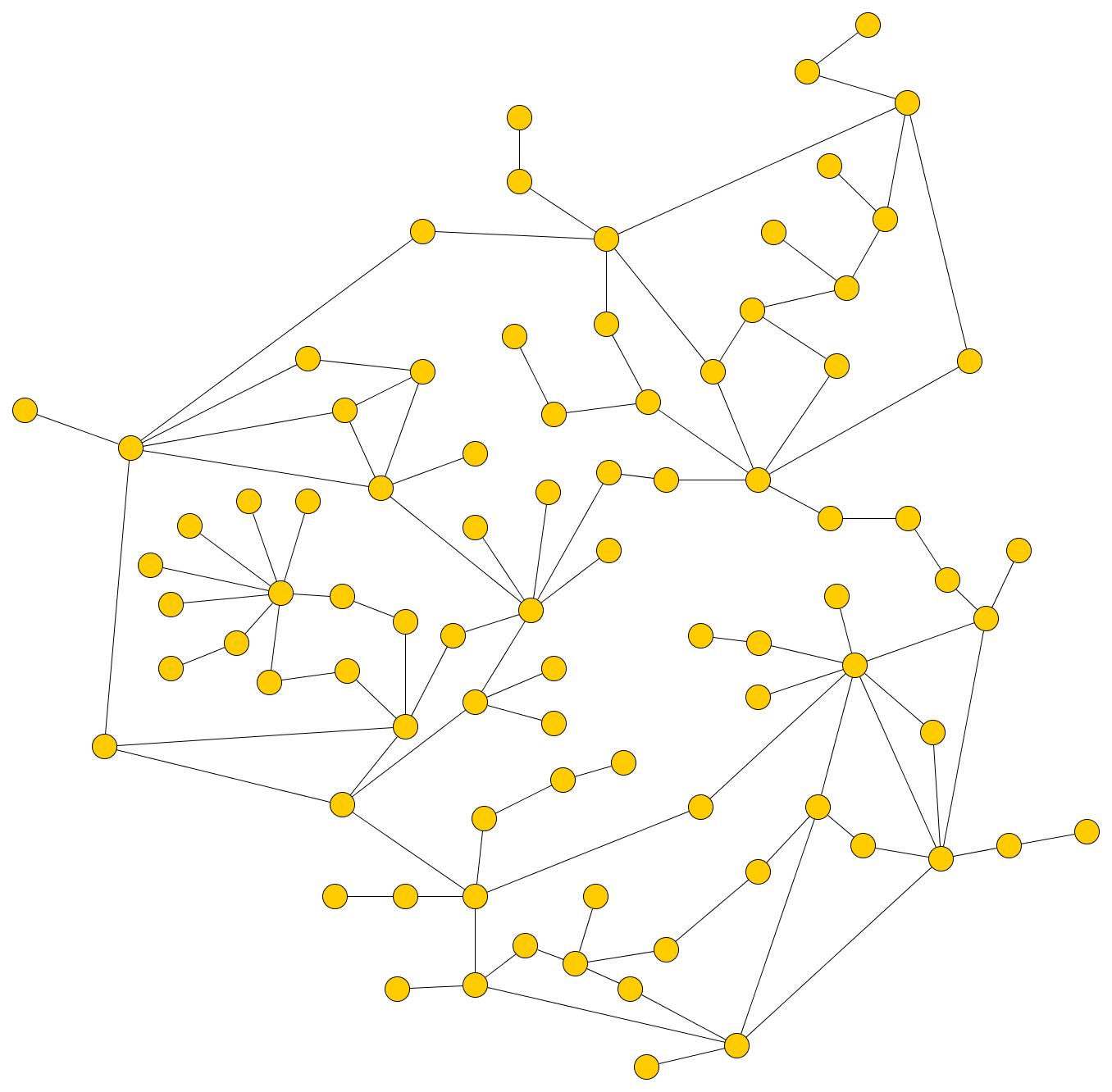}\\\midrule
     \multicolumn{6} {c} {\textbf {Fast Response Time (Seconds)}}\\\midrule
    & $G_5A_{\mathrm{FD}}$/$G_5D_3$ & $G_8A_O$/$G_8D_1$ & $G_5A_C$/$G_5D_4$ & $G_7A_O$/$G_7D_1$ & $G_9A_{\mathrm{MDS}}$/$G_9D_1$\\
     & ($16.86s$) & ($19.16s$) & ($19.44s$) & ($19.50s$) & ($19.98s$)\\
    & prop $=0.58$ & prop $=0.68$ & prop $=0.58$ & prop $=0.66$ & prop $=0.62$\\
     \begin{sideways}\textbf {Algorithm}\end{sideways} & \includegraphics[width=20mm]{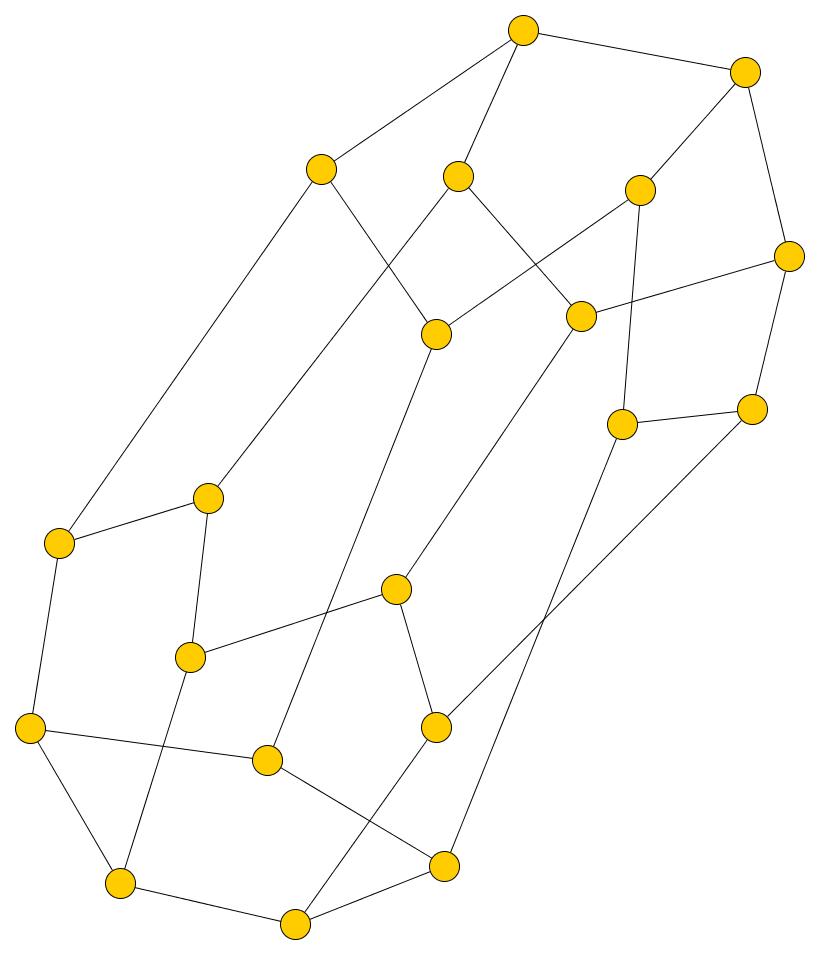} & \includegraphics[width=20mm]{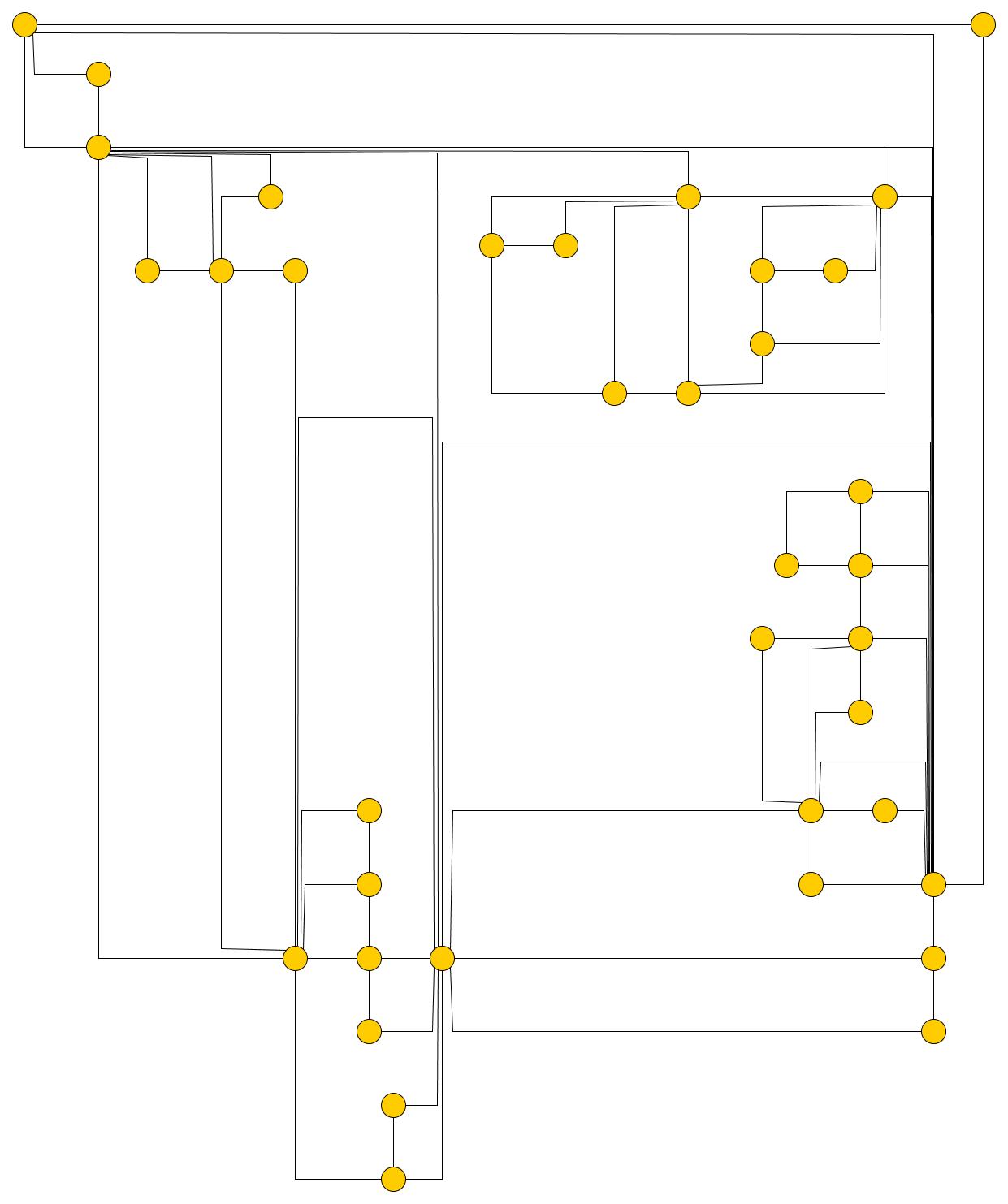} & \includegraphics[width=20mm]{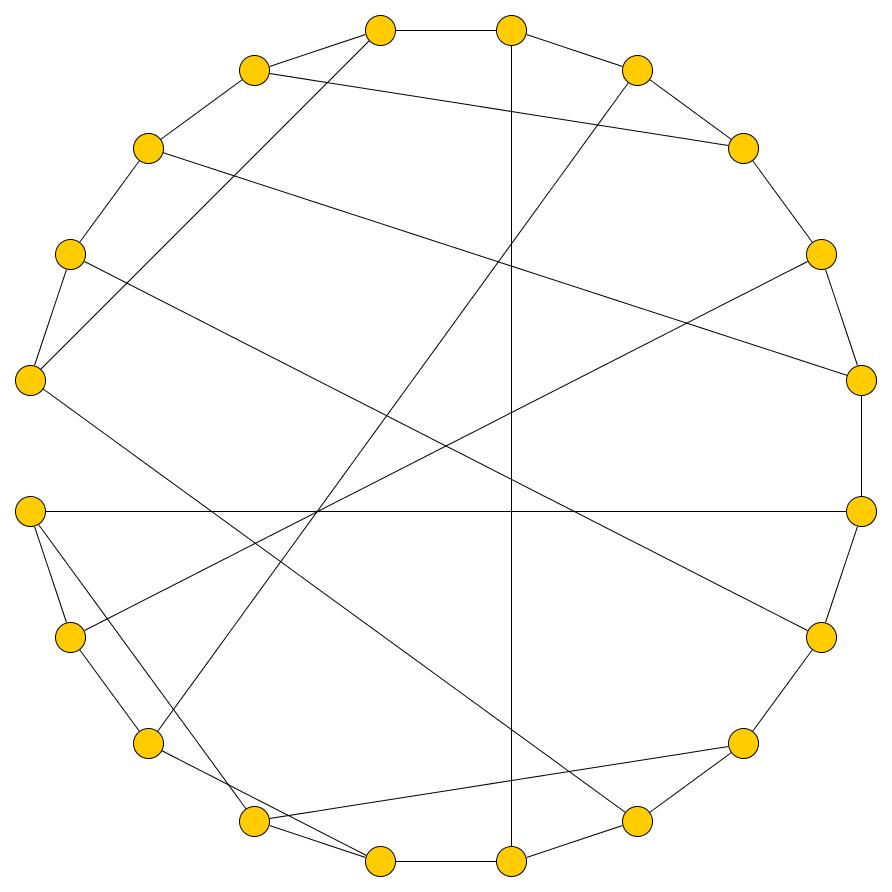} & \includegraphics[width=24mm]{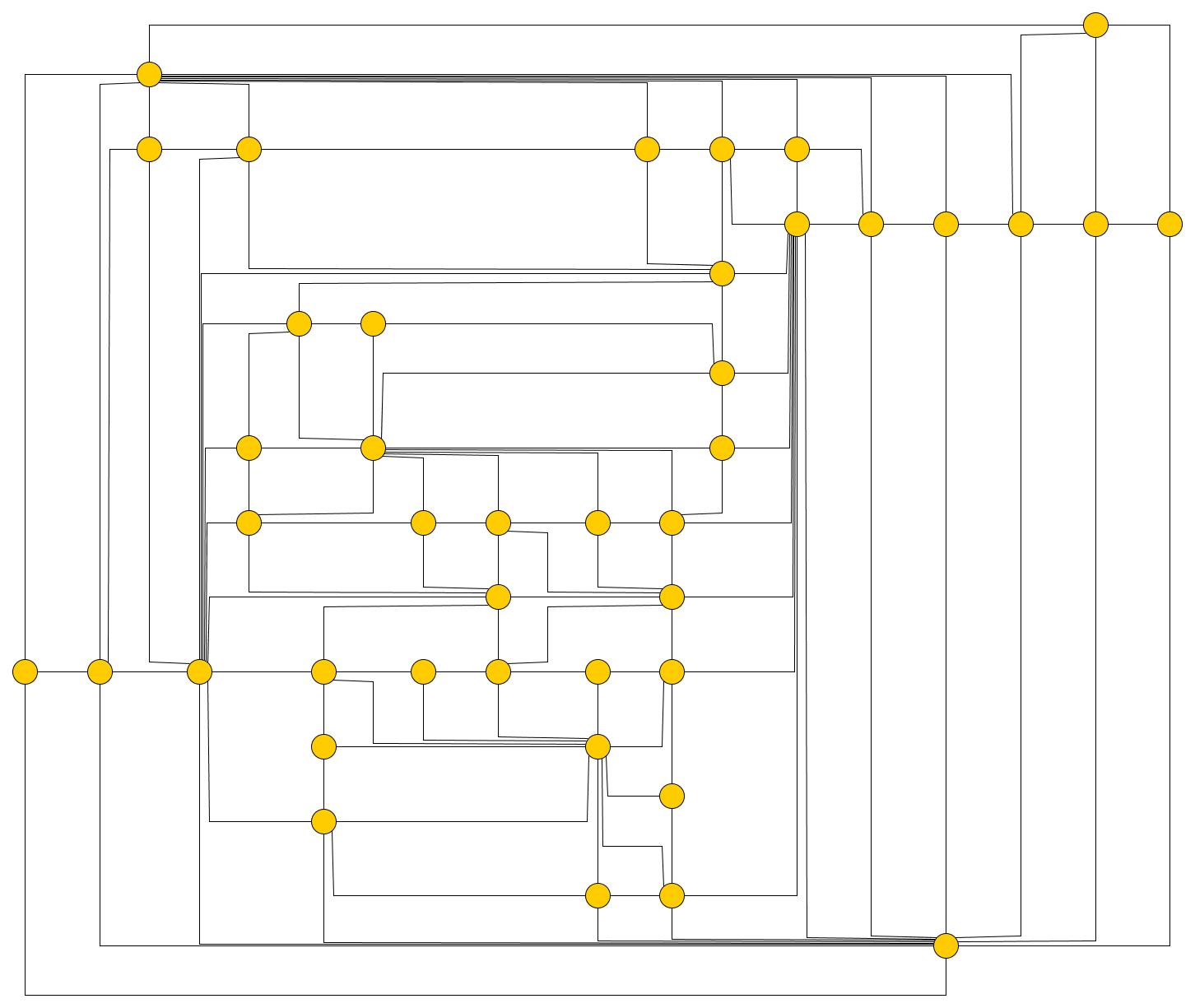} & \includegraphics[width=20mm]{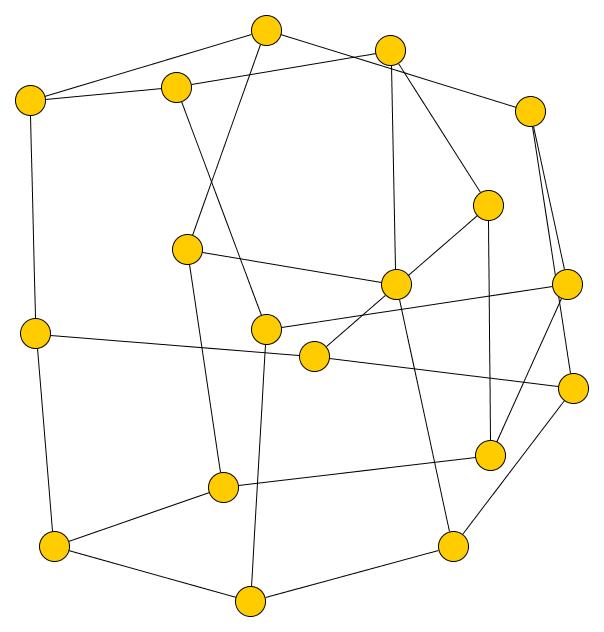}\\
     \begin{sideways}\textbf {Human}\end{sideways} & \includegraphics[width=20mm]{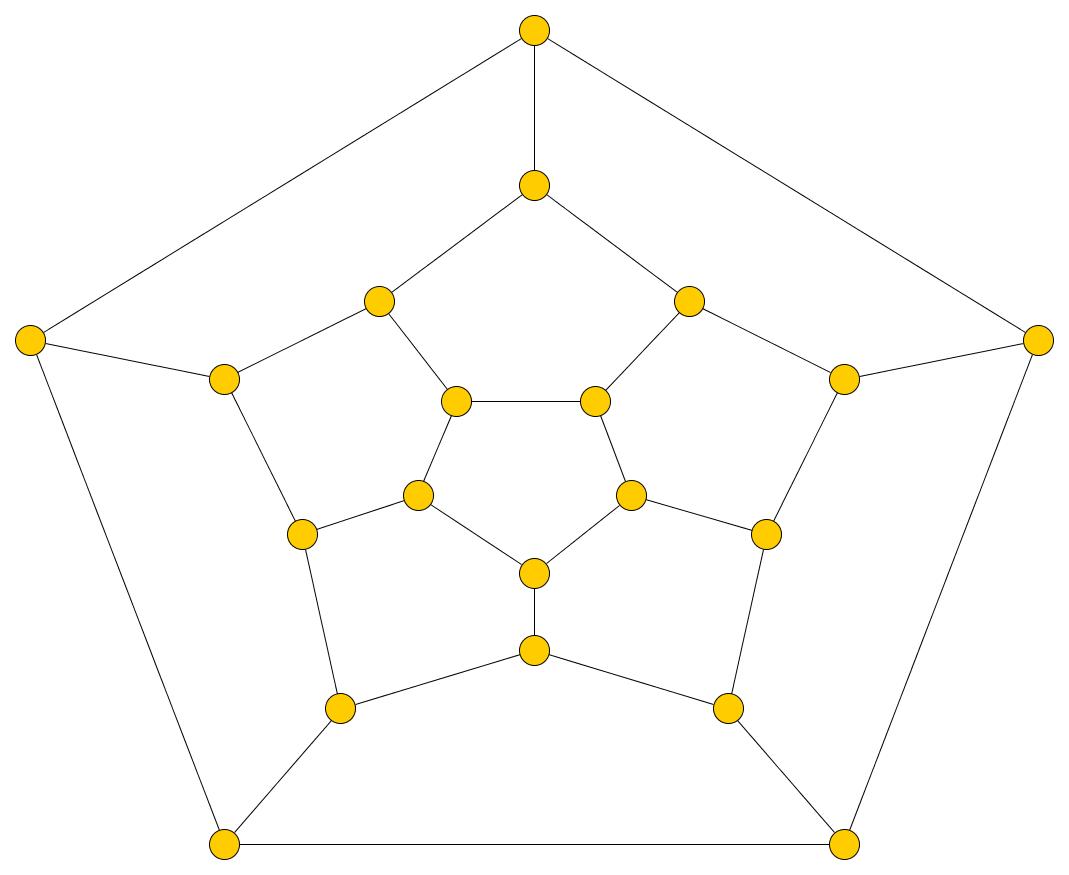} & \includegraphics[width=28mm]{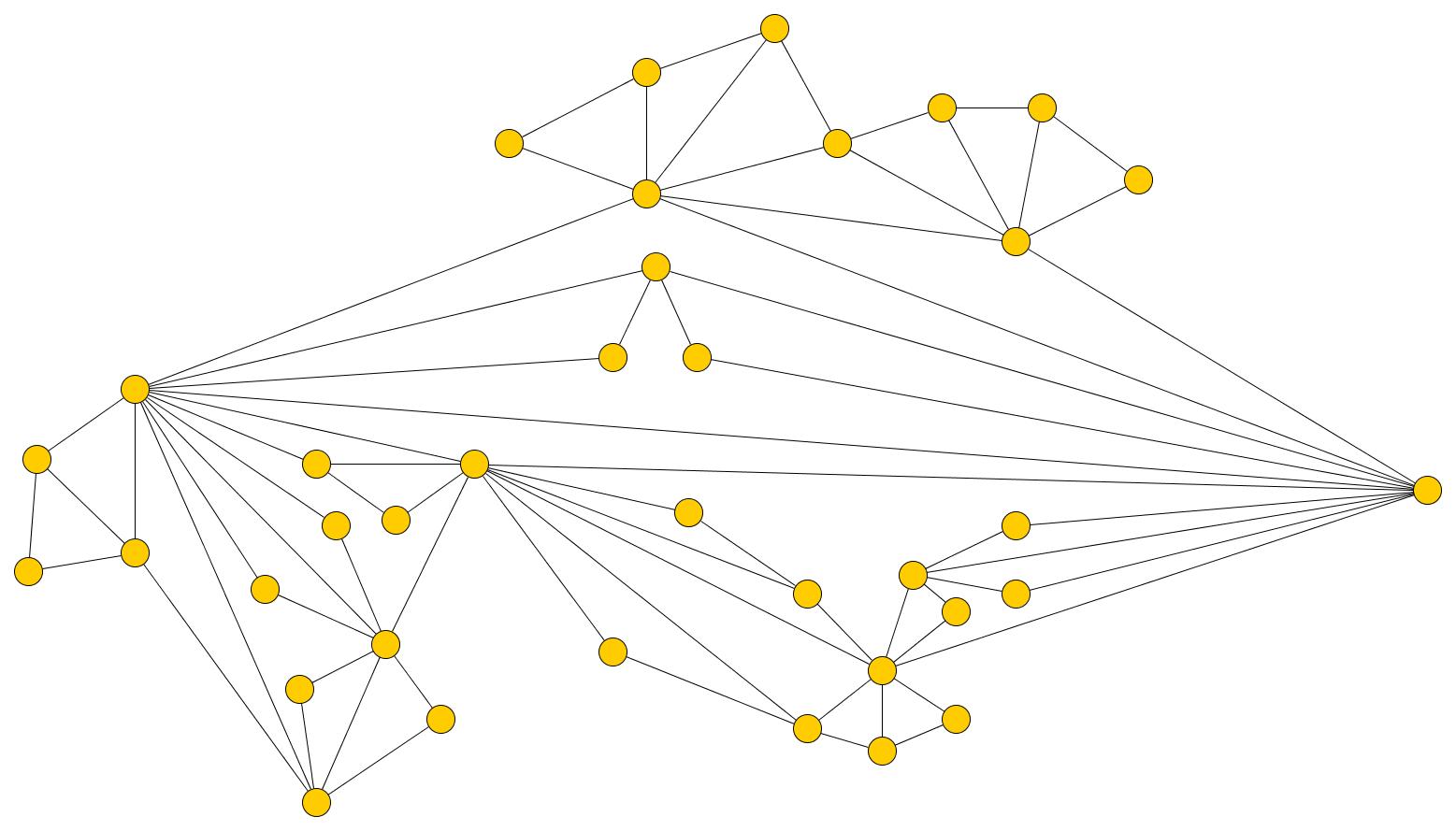} & \includegraphics[width=20mm]{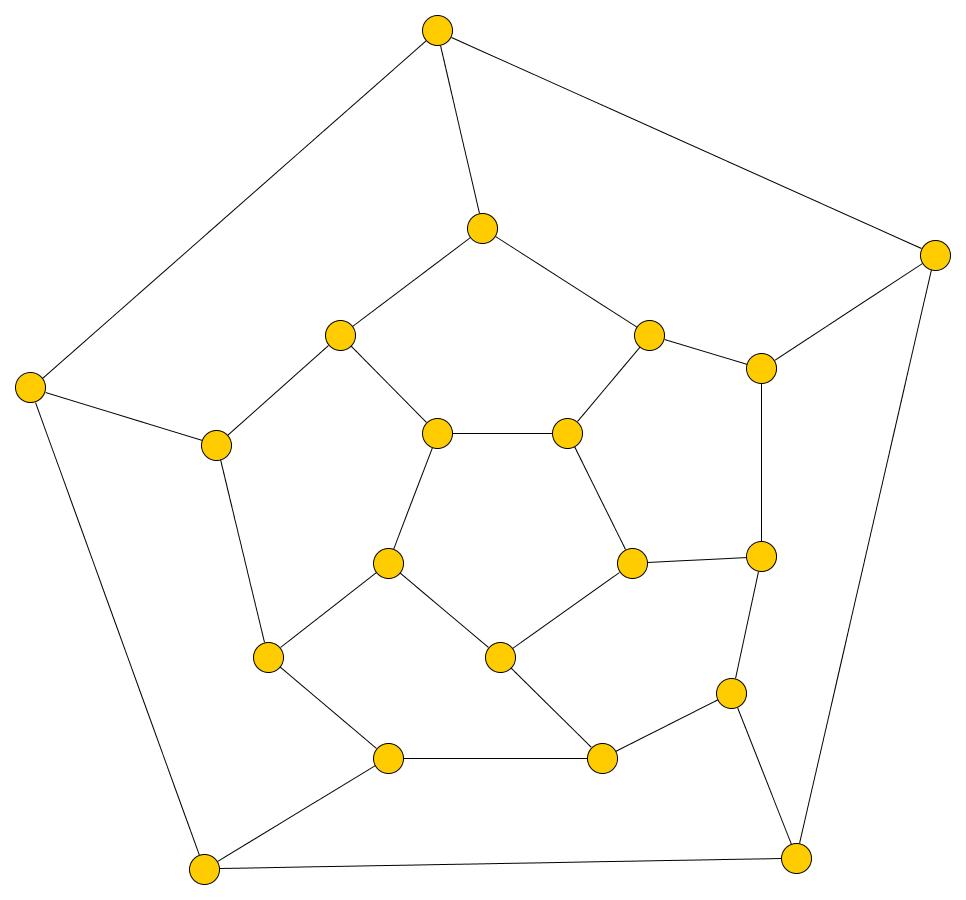} & \includegraphics[width=24mm]{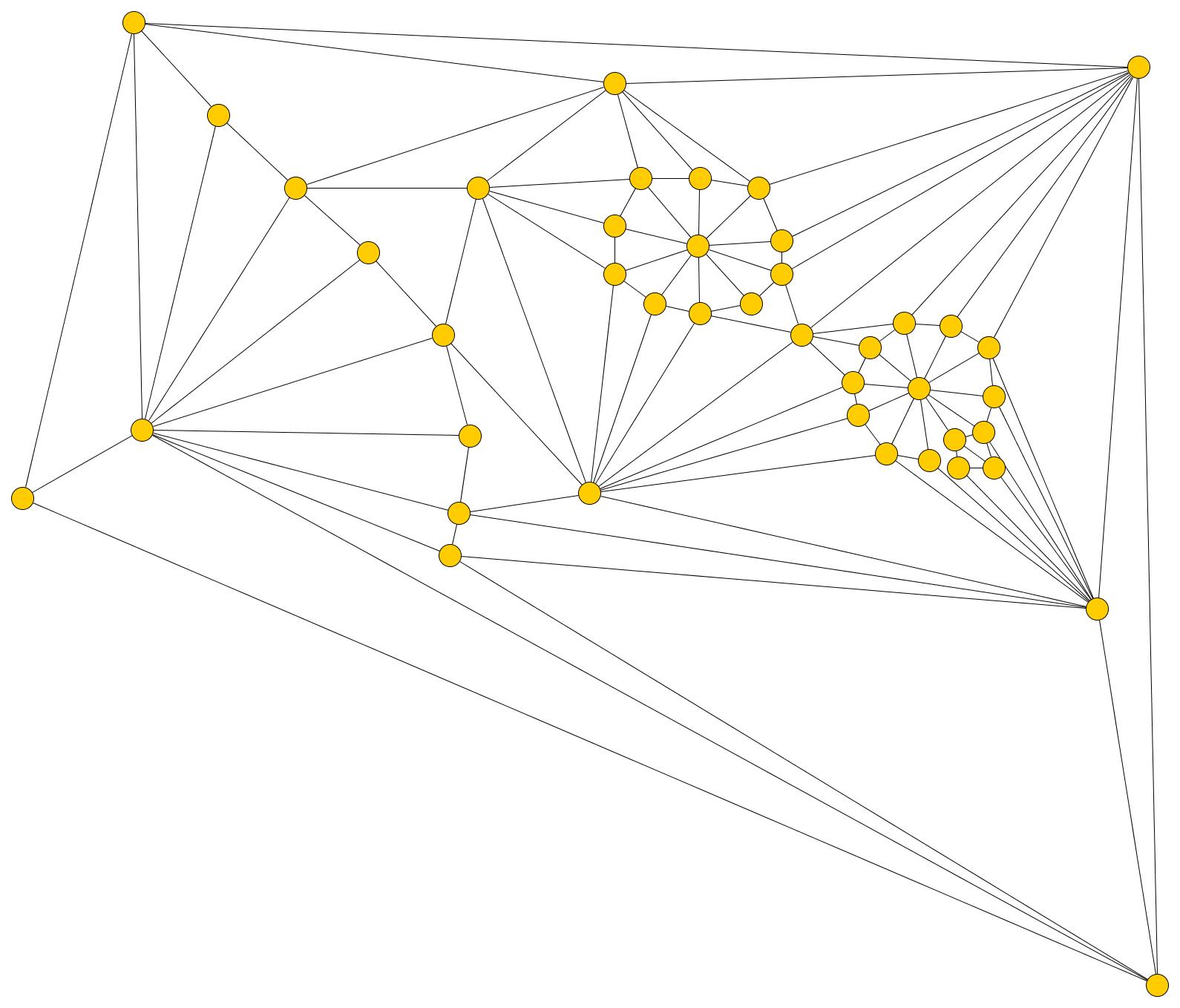} &\includegraphics[width=21mm]{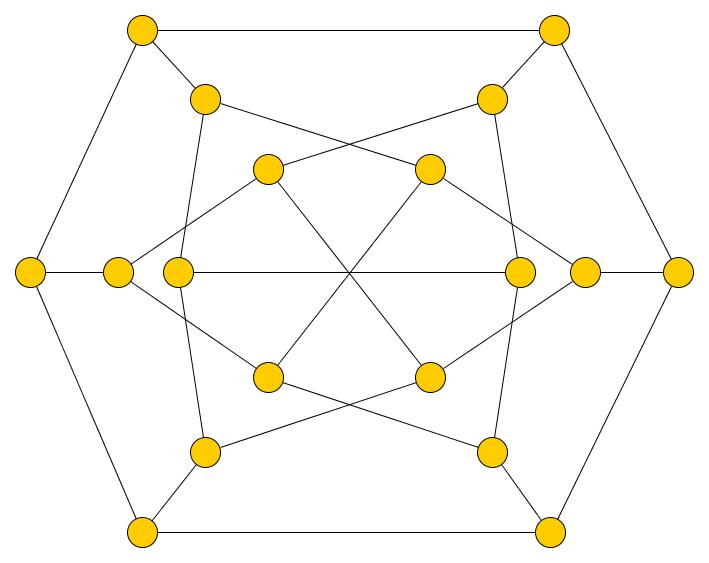}\\\bottomrule
  \end{tabular}
\caption {Trials with slow response times (top) and quick response times (bottom).  Time in seconds, and human-selection proportion shown.}
  \label{extremeTime}
\end{figure}

\begin{figure*}[t]
  \centering
  \scriptsize
\begin{tabular}{lcccc}\toprule
  \multicolumn{5}{c}{\textbf {High proportion of correct answers (human selected)}}\\\midrule
    & $G_6A_{\mathrm{FD}}$/$G_6D_3$ & $G_1A_C$/$G_1D_2$ & $G_1A_C$/$G_1D_1$ & $G_5A_{\mathrm{MDS}}$/$G_5D_2$\\
       & ($27.27s$) & ($37.89s$) & ($32.01s$) & ($32.40s$)\\
    & prop $=0.85$ & prop $=0.85$ & prop $=0.86$ & prop $=0.92$\\
    \begin {sideways}\textbf {Algorithm} \end{sideways} & \includegraphics[width=26mm]{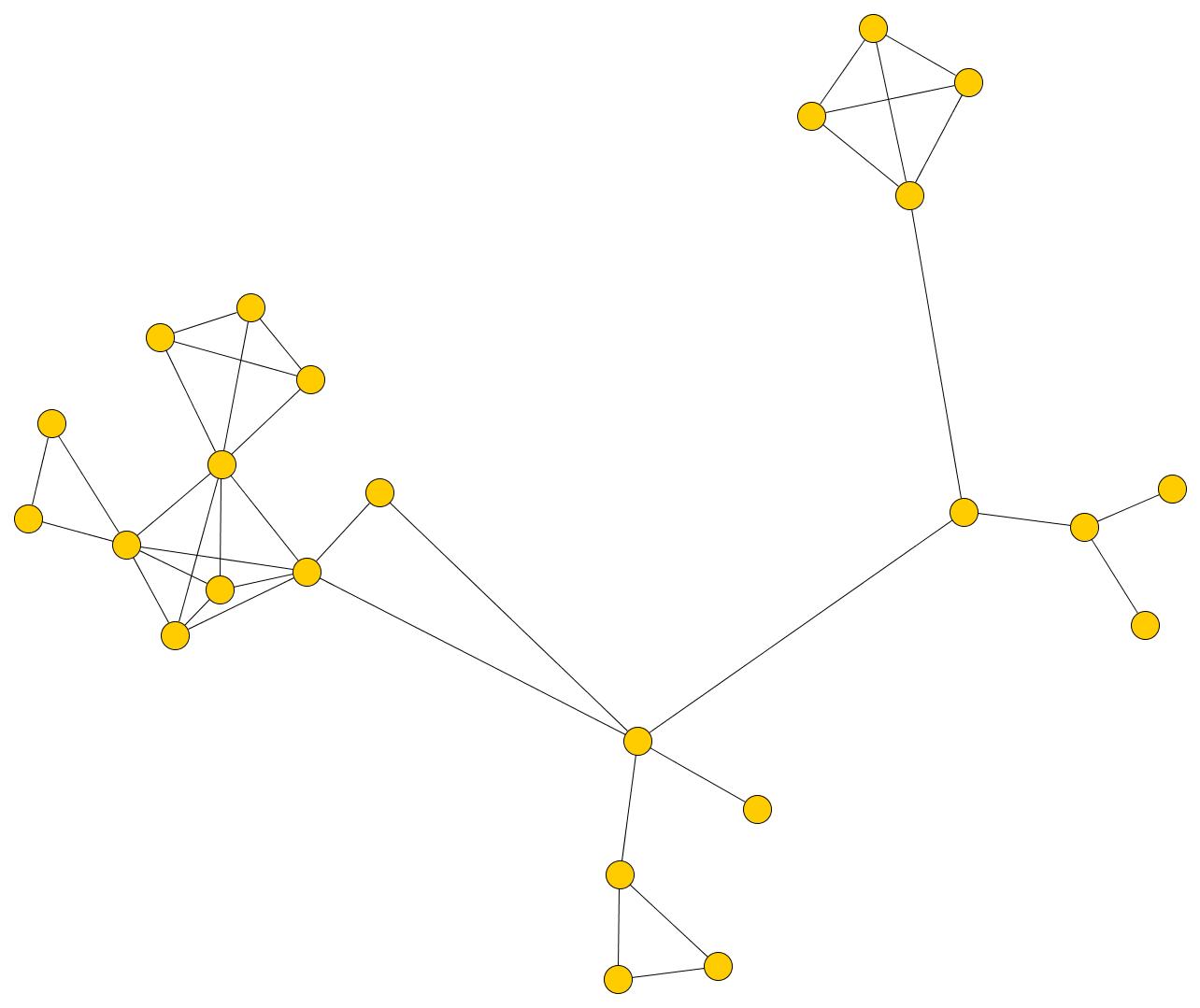} & \includegraphics[width=25mm]{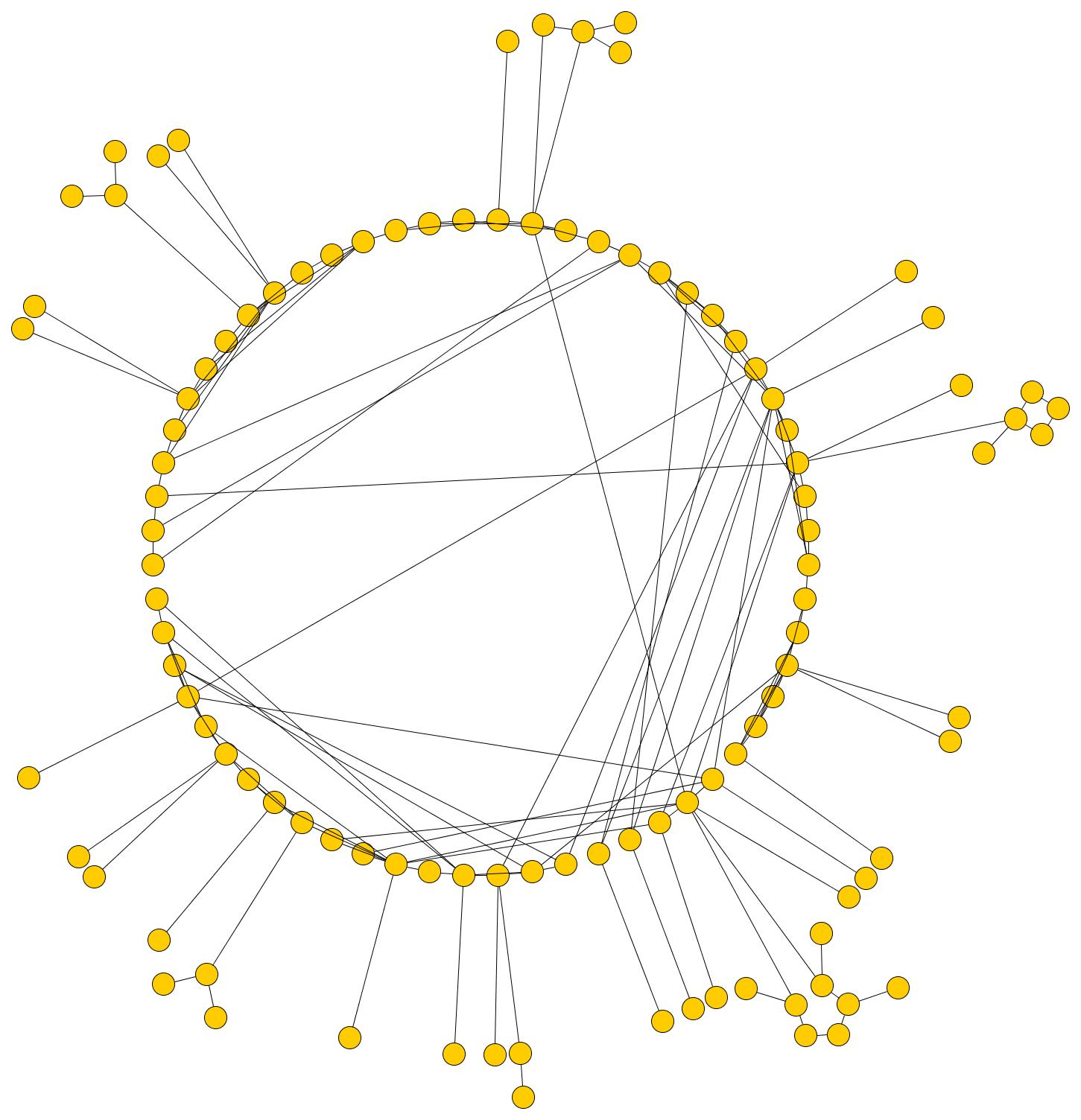} & \includegraphics[width=25mm]{figures/g1a3} & \includegraphics[width=24mm]{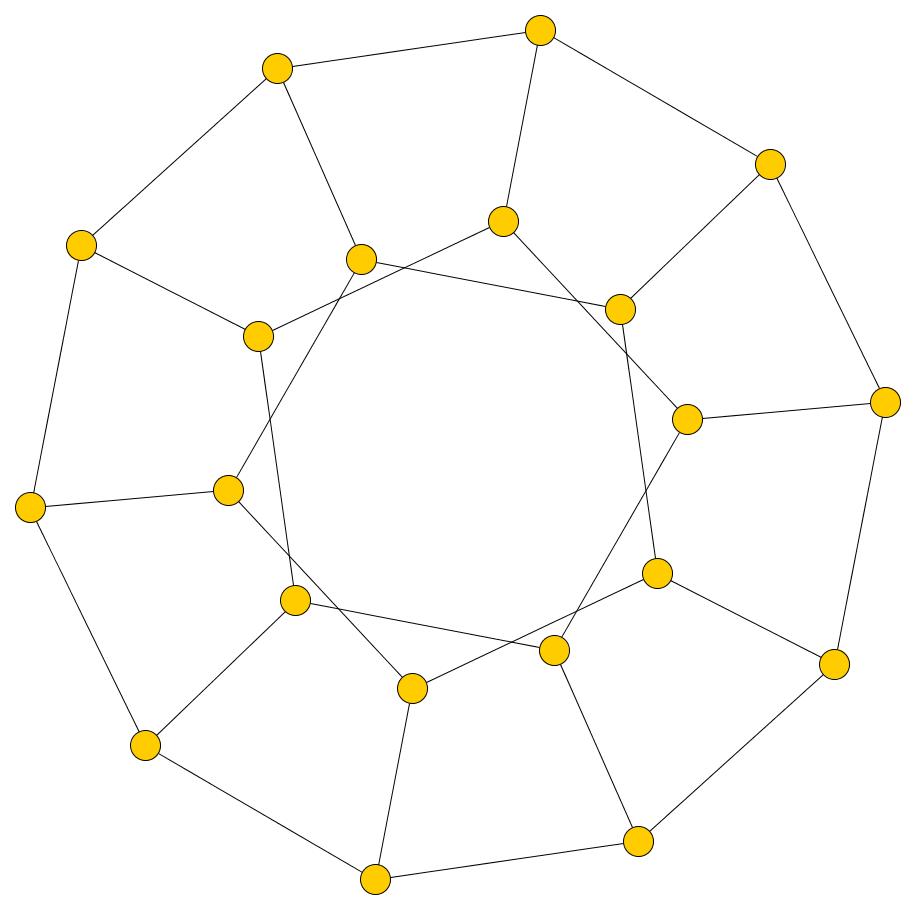}\\
    \begin{sideways}\textbf {Human}\end{sideways} & \includegraphics[width=18mm]{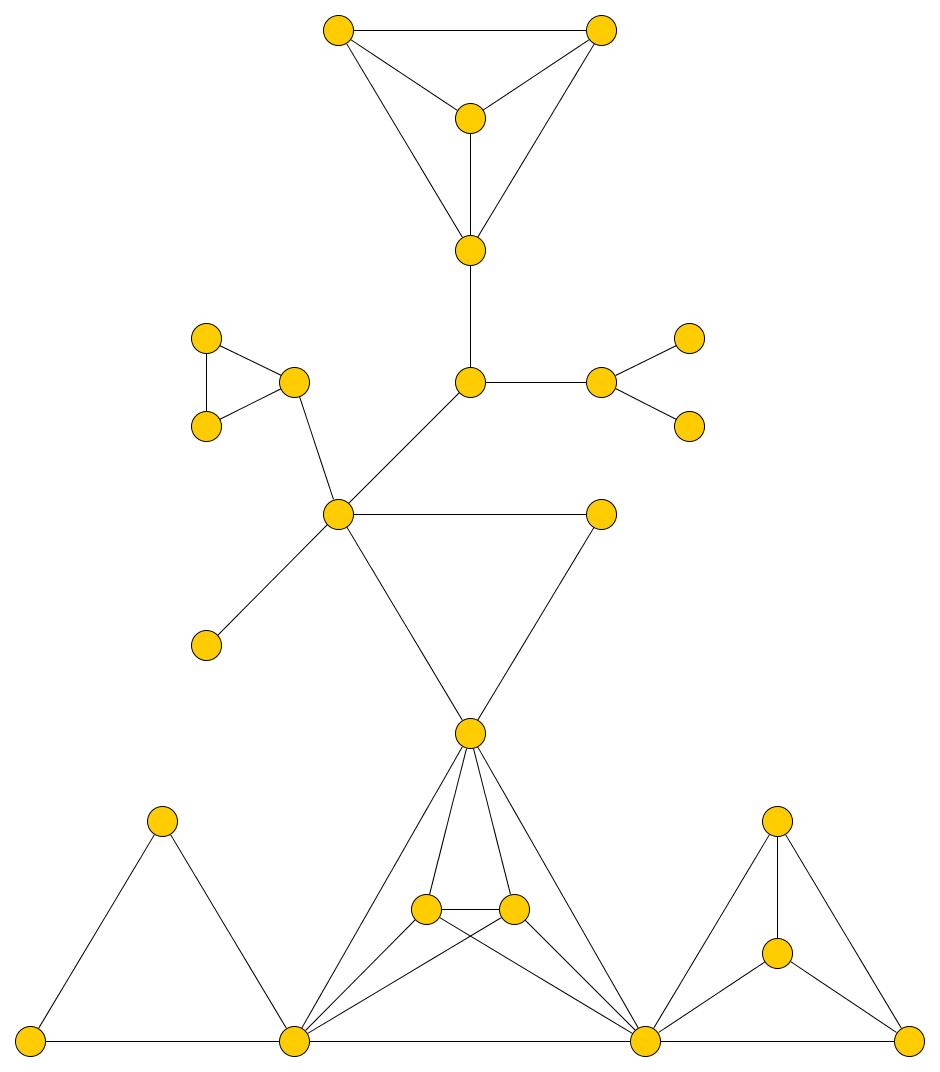} & \includegraphics[width=23mm]{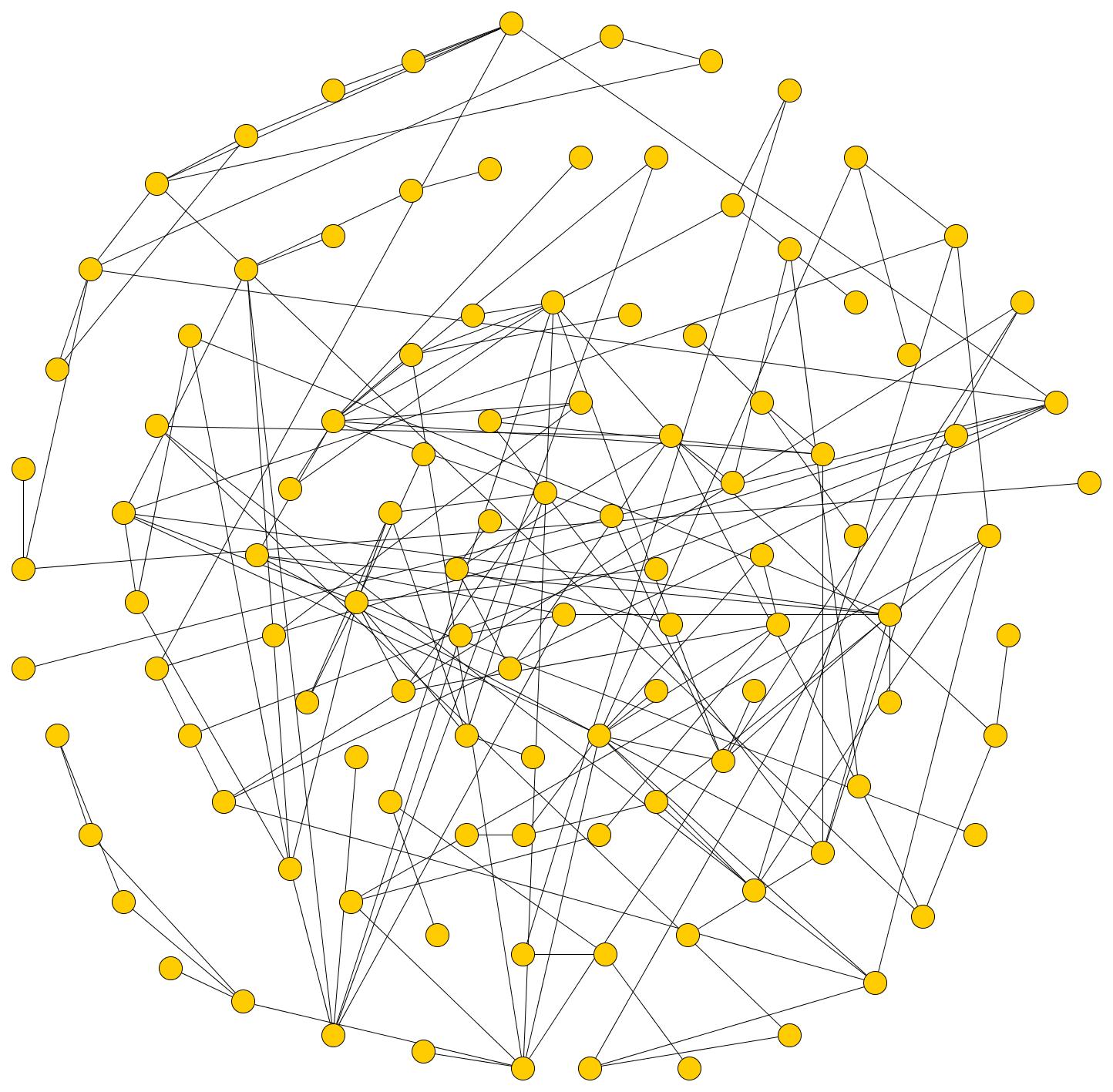} & \includegraphics[width=25mm]{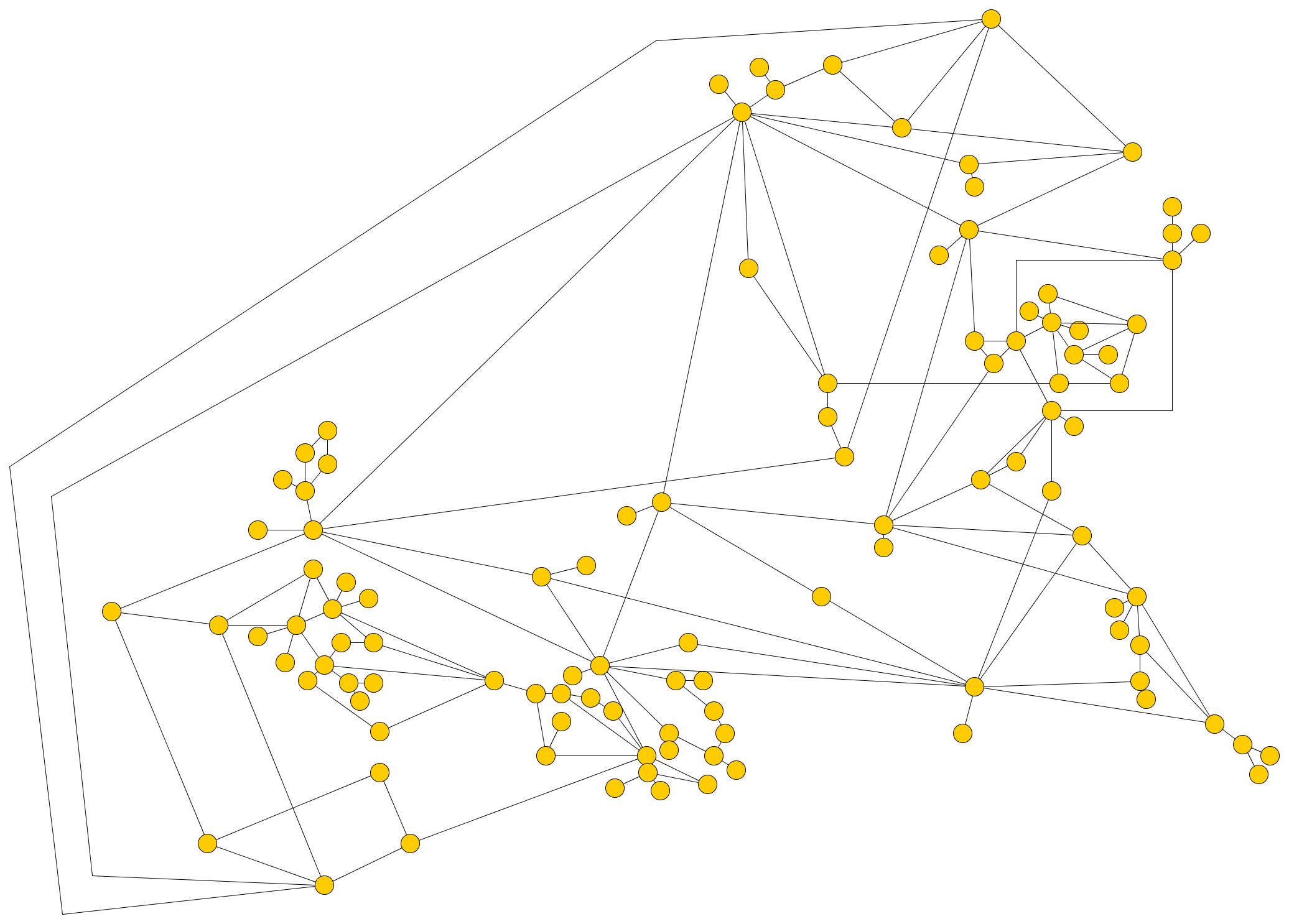} & \includegraphics[width=24mm]{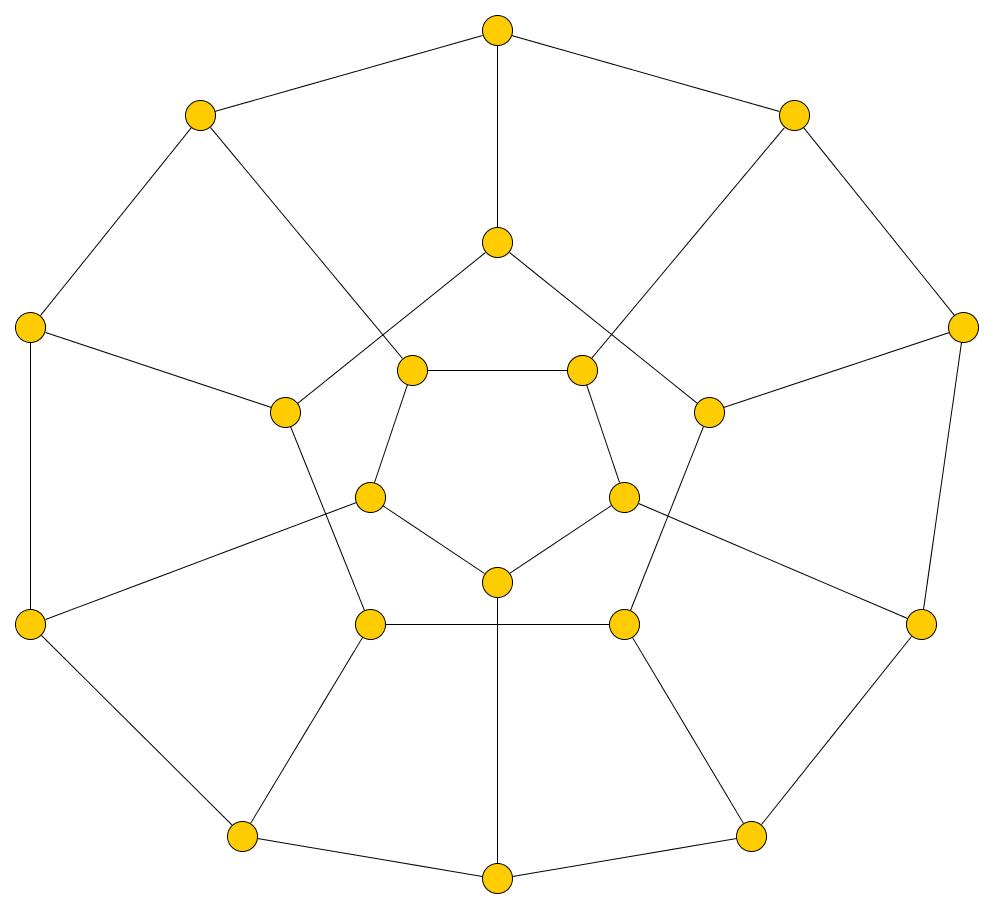}\\\bottomrule
\end {tabular}

  \scriptsize
\begin{tabular}{lccccc}\toprule
    \multicolumn{6} {c} {\textbf {High proportion of incorrect answers (algorithm selected)}}\\\midrule
    & $G_3A_{\mathrm{MDS}}$/$G_3D_1$ & $G_9A_C$/$G_9D_2$ ($28.6s$) & $G_2A_{\mathrm{MDS}}$/$G_2D_1$ & $G_5A_{\mathrm{FD}}$/$G_5D_1$ & $G_3A_{\mathrm{MDS}}$/$G_3D_3$ \\
     & ($38.59s$) & ($28.6s$) & ($23.14s$) & ($26.60s$) & ($34.30s$)\\
    & prop $=0.18$ & prop $=0.23$ & prop $=0.27$ & prop $=0.27$ & prop $=0.29$\\
     \begin{sideways}\textbf {Algorithm}\end{sideways} & \includegraphics[width=20mm]{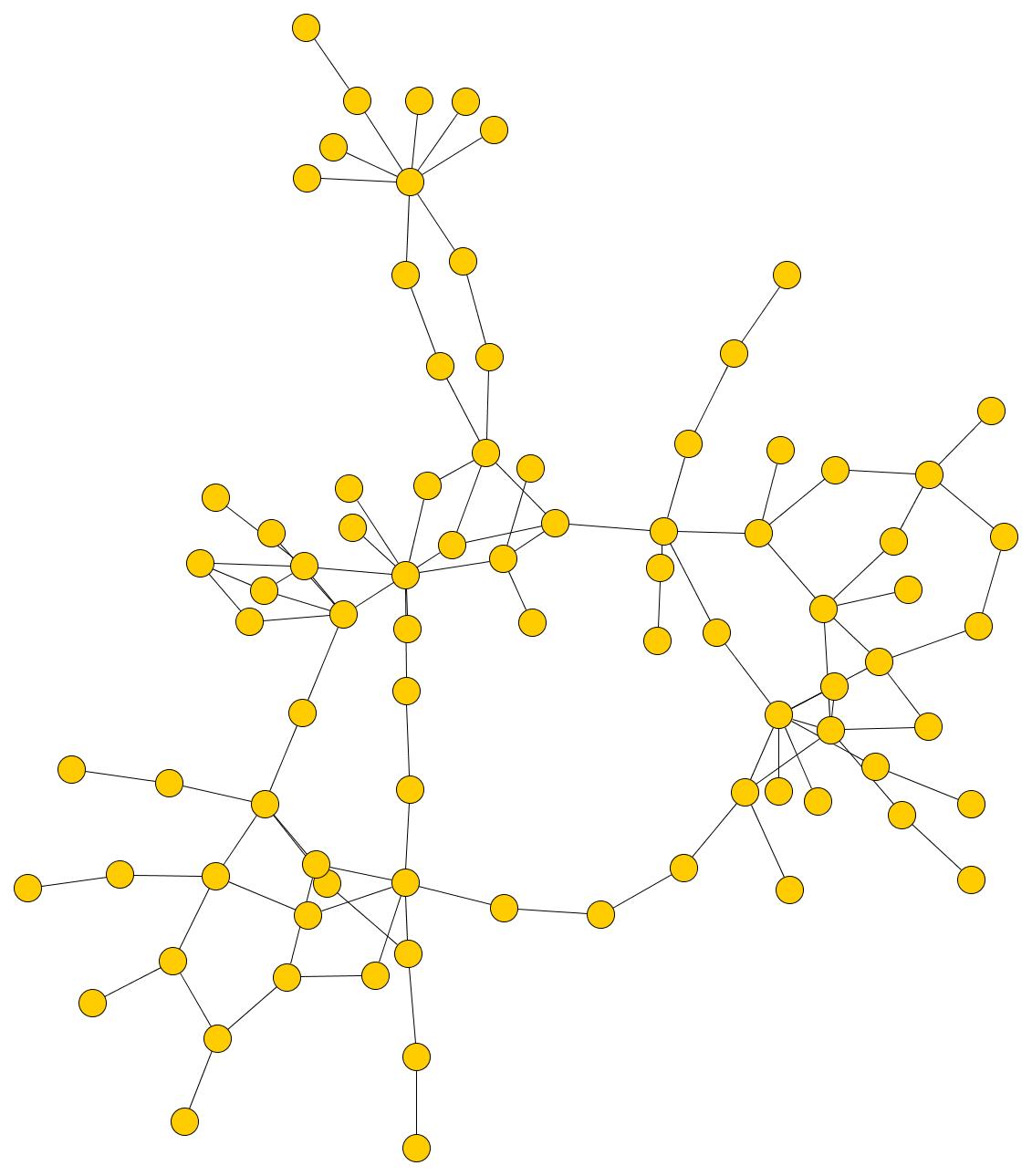} & \includegraphics[width=20mm]{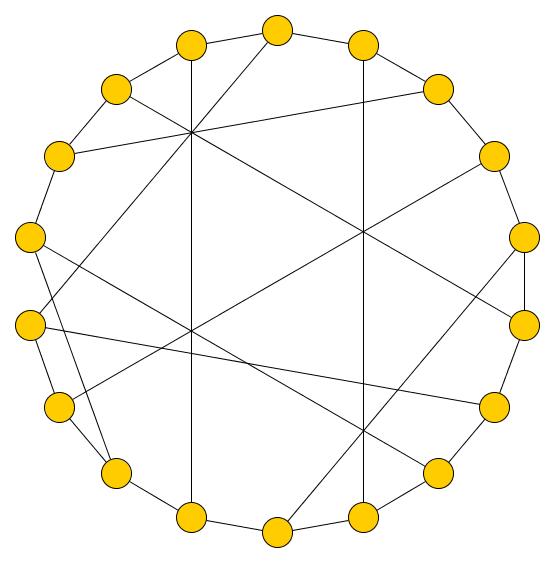} & \includegraphics[width=20mm]{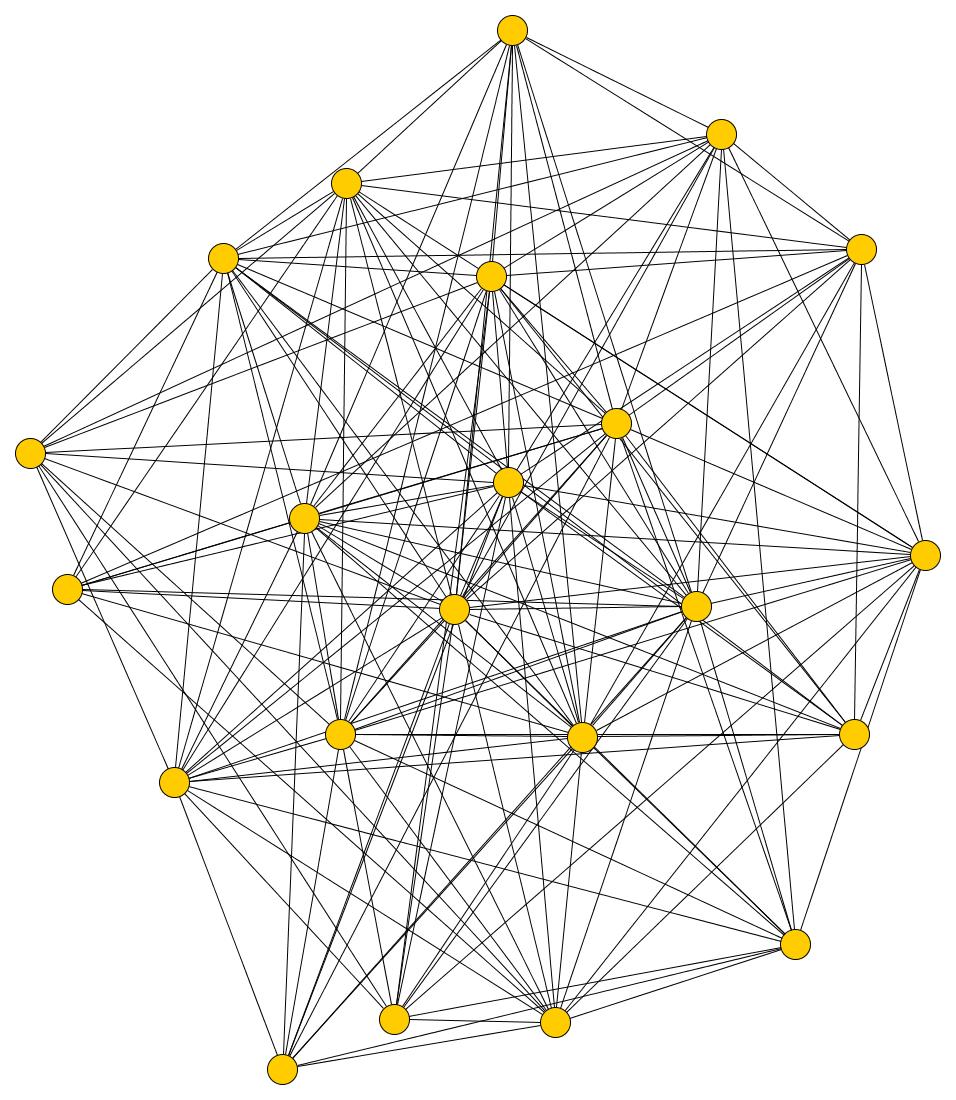} & \includegraphics[width=20mm]{figures/g5a1} & \includegraphics[width=20mm]{figures/g3a2}\\
     \begin{sideways}\textbf {Human}\end{sideways} & \includegraphics[width=24mm]{figures/g3h2} & \includegraphics[width=20mm]{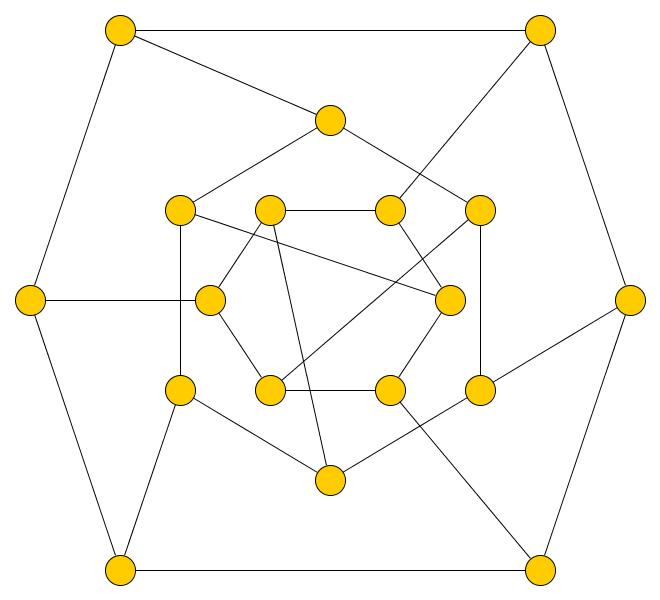} & \includegraphics[width=24mm]{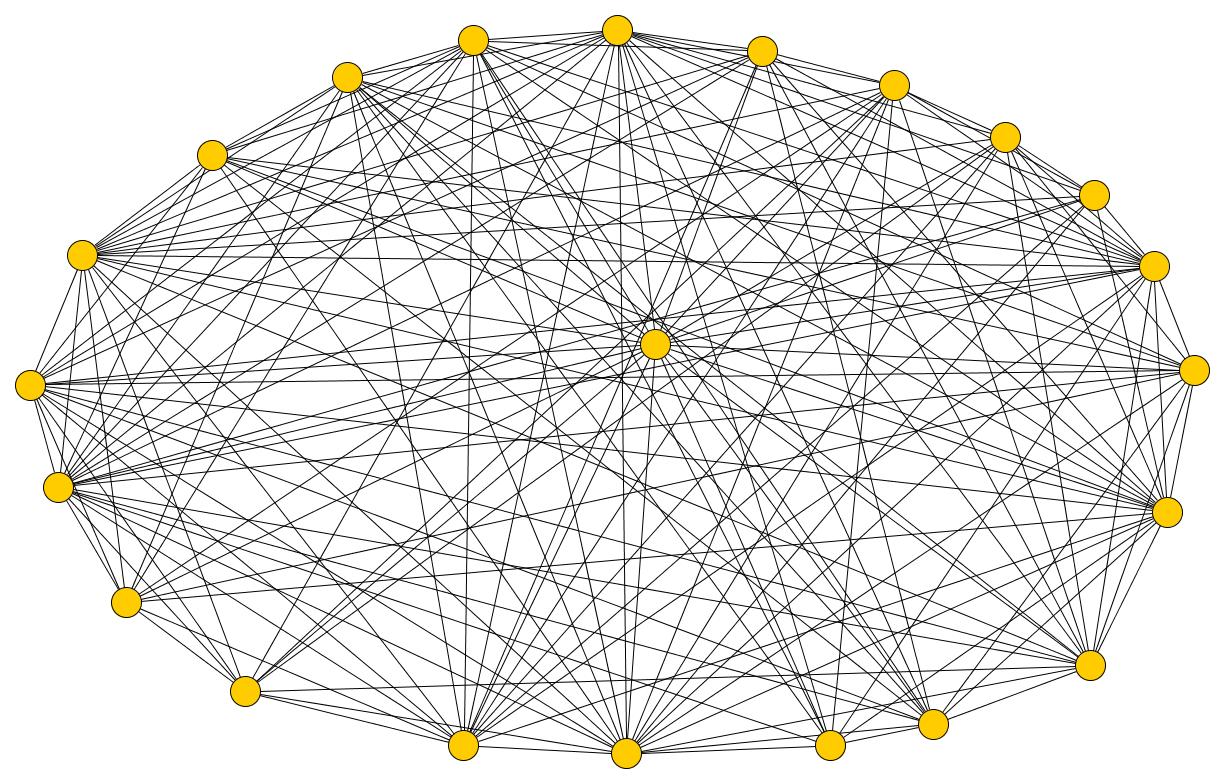} & \includegraphics[width=16mm]{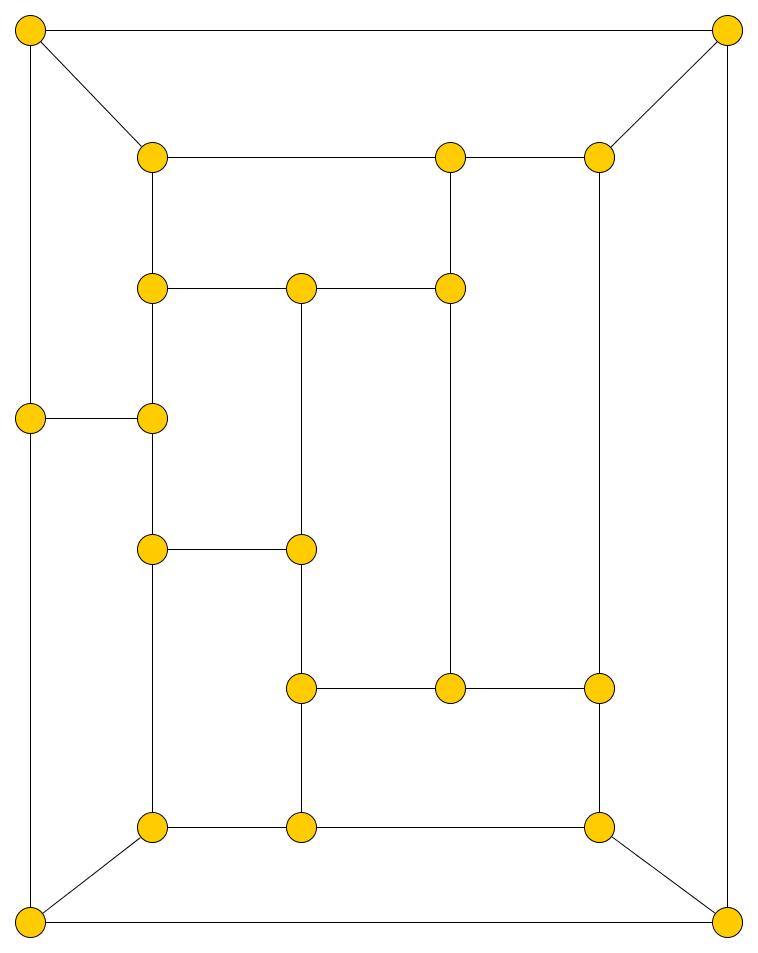} &\includegraphics[width=20mm]{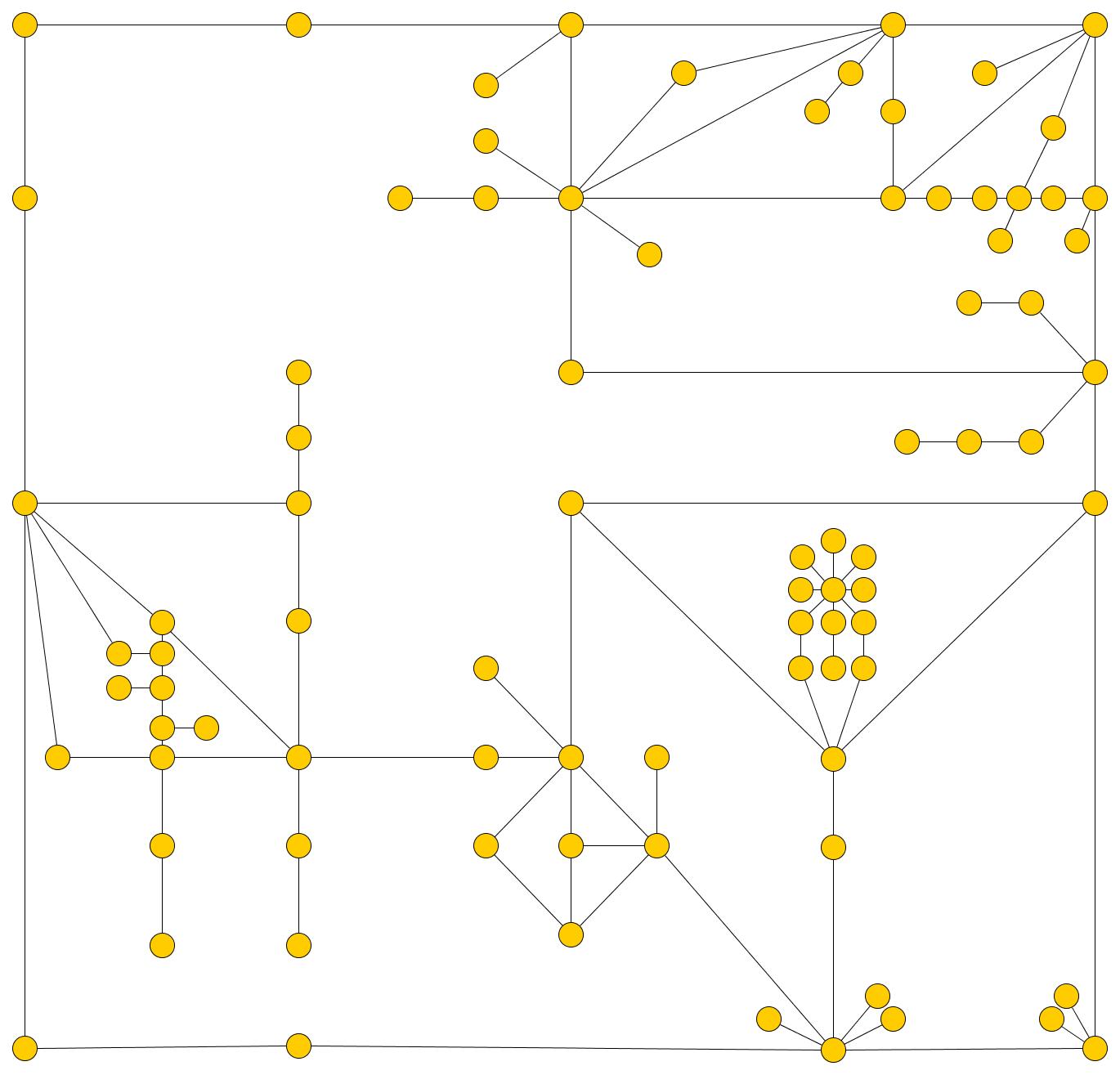}\\\bottomrule
\end {tabular}
  \caption {Trials with a high proportion of correct (human drawing chosen, upper) and incorrect (algorithm drawing chosen, lower) answers.}
  \label{extremeProp}
\end{figure*}

Extreme trials (response time: Figure~\ref{extremeTime}; proportion: Figure~\ref{extremeProp}) are identified as $G_iA_j$ and $G_iD_k$: $G_i$ (graph), $A_j$ (algorithm), $D_k $(drawer). All experimental stimuli jpeg files can be found in the supplementary material (visit \url{http://www.dcs.gla.ac.uk/~hcp/GD2020}).

Three slow trials relate to a particular FD graph, suggesting that this form of drawing was seen by participants as possibly hand-drawn -- it shows clusters and symmetry, while the drawers all attempted to remove crosses. The combinations of $G_4A_{\mathrm{MDS}}$/$G_4D_4$ and $G_7A_C$/$G_7D_4$ (top row of Figure~\ref{extremeTime}) are interesting because, for each, the overall shape of the human-drawn graph is similar to that produced by the algorithm: it is not hard to see why participants found this choice difficult. Three quick responses ($G_5A_{\mathrm{FD}}$/$G_5D_3$, $G_5A_C$/$G_5D_4$, $G_9A_{\mathrm{MDS}}$/$G_9D_1$, bottom row of Figure~\ref{extremeTime})  demonstrate effort on the part of the drawer to depict symmetry that is not highlighted by the algorithms; the other two relate to the orthogonal algorithm, which, as noted above, produced worst performance in making a human {\it vs} algorithm judgements.

Of the four combinations where participants gave mostly correct responses, it is not hard to see why for $G_1A_C$/$G_1D_2$ and $G_1A_C$/$G_1D_1$ (top row of Figure~\ref{extremeProp}), since the human-drawn graphs lack any clear structure or visual elegance in comparison with those created by the circular algorithm. The fact that $G_5A_{\mathrm{MDS}}$ is geometrically precise in its node positioning (while $G_5D_2$ has slight mis-positionings) can explain the 0.92 accuracy for this combination, although we note that this decision still took above average time (32.4 seconds). More difficult to explain is the high proportion associated with $G_6A_{\mathrm{FD}}$/$G_6D_3$, since the human drawing is highly structured and symmetrical. Of the combinations where the average accuracy is low, three  algorithmic drawings depict some extent of symmetry ($G_3A_{\mathrm{MDS}}$, $G_9A_C$, $G_5A_{\mathrm{FD}}$, bottom row of Figure~\ref{extremeProp}), while the fourth is compared against a human drawing which used an approach that, if adopted by an algorithm, would have resulted in a more geometrically precise diagram. The examples in Figure~\ref{extremeProp} (top and bottom rows) suggest that regular node and edge placements (that is, grid-like or evenly spaced on a circle), indicate an algorithmically-drawn graph.

Key factors affecting the human {\it vs} algorithm choice were thus depiction of symmetry (even if only approximate), and geometric precision (i.e.~very precise node placement, with regular spacing or grid-like).

\section{Discussion}

In general, over all graphs and algorithms, participants can correctly distinguish hand-drawn layouts from algorithmically created ones: graph drawing algorithms (in general) effectively fail the Turing Test. The only exception is the Force-Directed algorithm, where we did not find evidence that participants could reliably distinguish between the algorithmic and hand-drawn layouts. We speculate this might be because our drawers (consciously or unconsciously) created drawings with similar FD layout principles in mind: separating unconnected nodes, and clustering connected ones together. The MDS algorithm provided some evidence of passing the test (in particular for medium and large graphs); it produces similar shapes to FD.

We were not surprised that it was easy to distinguish circular (especially large circular) and orthogonal graph drawings from hand-drawn ones, since they make use of precise node placement: equal separation around the circle circumference, placement on equally-spaced horizontal lines or on an underlying unit grid. While the human drawers sometimes used such placements ($G_2D_1$ and $G_5D_1$ in Figure~\ref{extremeProp}), in many cases ($G_8D_1$ in Figure~\ref{extremeTime}, $G_5D_2$ in Figure~\ref{extremeProp}) they did not. We were also not surprised to find that larger graphs took more time than the smaller ones, but were surprised that experts took longer than novices -- we had expected the converse; perhaps experts made more considered analytical decisions as opposed to novices' more spontaneous ones.

\section{Subjective Quality of the Drawings}

Our study shows that some graph drawing algorithms produce diagrams that are obviously perceived as different from those drawn by graph drawing experts. This raises the question: if algorithmic drawings are perceived as being different from hand-drawn ones, are they any better? And even if they are not perceived as different, is there a perceived difference in quality?

We followed our Turing experiment with a supplementary, almost identical study, using the same paired stimuli and experimental system. The only difference was the question asked: `Which drawing is better?'. We deliberately did not give a definition for `better', since (at least for this initial study), we wished to get an overall judgement, rather than, for example, one based on a particular task or defined aesthetic. 52 participants took part, producing a total of 4887 data points. As before, hand-drawn graphs are scored $1$, and algorithmic drawings $0$. Thus, proportions $> 0.5$ indicate the human drawing was, on average, considered better. Over all graphs and algorithms, the vote was for hand-drawn graphs (proportion=0.57, $p<0.001$). However, size and algorithm data show variations within this overall result (Table~\ref{sizeAlgWhichBetter}). Hand-drawn graphs were always preferred over orthogonal drawings; FD and MDS were preferred for medium and large graphs, and circular only for the large graphs.

\begin{table*}[t]
\caption {Results for the `Which is better' question, by graph size and algorithm. * indicates statistically significant results ($p<0.05/12 = 0.0042$)}
\centering
\scriptsize
\begin{tabular}{llllllllll} \toprule
  &\multicolumn{2}{c}{Force-Directed} &\multicolumn{2}{c}{MDS} &  \multicolumn{2}{c}{Circular} &  \multicolumn{2}{c}{Orthogonal}\\
  &proportion &p-value &proportion &p-value&proportion &p-value&proportion &p-value\\\midrule
  small & 0.83* & $<0.001$ & 0.68* & $<0.001$ & 0.55 & 0.040 & 0.62* & $<0.001$\\
  medium & 0.44 & 0.006 & 0.42* & 0.001 & 0.62* & $<0.001$ & 0.74* & $<0.001$\\
  large & 0.19* & $<0.001$ & 0.42 & 0.009 & 0.41* & 0.002 & 0.63* & $<0.001$\\\bottomrule
\end{tabular}
\label {sizeAlgWhichBetter}
\end {table*}

Thus, even when hand-drawn and algorithmic drawings are indistinguishable (as shown for FD and MDS in the first experiment), subjective judgement (experiment 2) determines that the algorithmic versions are `better', especially for the larger graphs. The orthogonal algorithm had no wins: it did not pass the Turing Test, and was always considered worse than the hand-drawn versions. There were mixed results for the circular algorithm: easy to distinguish from hand-drawn layouts when small or medium, and only preferred when large.

\section{Conclusions and Future Work}

This is the first experiment that compares graphs drawn by graph drawing researchers to those produced by graph drawing algorithms as a Turing Test. Overall, we found that hand-drawn graphs could be reliably distinguished from those generated by algorithms -- thus, on average, Turing Test failure. However, we did not find evidence that force-directed and (marginally) MDS algorithms could be reliably distinguished from hand-drawn layouts -- they therefore effectively `pass' the Turing Test. We speculate that this is the case because of the prevalence of these algorithms in the popular media (e.g., for depicting social networks); further studies could establish exactly why these two algorithms perform differently from the others. 

The generalisability of our conclusions is, of course, limited by our experimental scope. While we used a good range of real-world and abstract graphs, differently sized graphs, planar and non-planar graphs, and good coverage of various graph metrics, our data set comprises nine experimental graphs. Using only `small' graphs (15 to 108 nodes) was an obvious design decision when considering the feasibility of creating hand-drawn layouts. We chose four common layout algorithms representing different approaches, and four human drawers (experts in graph drawing research). Despite these experimental limitations, our results represent the first empirical attempt to compare perception of a range of hand-drawn versus algorithmic graph layouts as a `Turing Test'.

Our motivation for these studies arose from a desire to determine whether algorithms depicting small graphs produce results that are similar to human efforts. Our results show that, in general, people notice when a graph has been hand-drawn. This result must, of course, be weighed against the length of time that it takes to draw a graph: we found that it takes much longer than we had anticipated to create drawings by hand. We also need to consider that, when considering the algorithmic approaches separately, some algorithmic versions were considered `better' than the hand-drawn ones -- the notable exception being the orthogonal algorithm.

Graph drawing algorithms are often inspired by assumptions about what a human would do in generating a drawing. Therefore, understanding what makes a drawing human-like will help inform future algorithm designers to make algorithms of higher quality. In future work, we would like to explore whether we get similar results if we explicitly match graph structure with graph algorithm (e.g., tree algorithms for trees, planar algorithms for planar graphs), use other less common algorithms (e.g., HOLA~\cite{kdmw-hhonl-16}, Wang et al.~\cite{Wang::2018:TVCG}), and use graphs drawn by a wider range of people (including non-experts). In addition, gathering both quantitative and qualitative data in future studies will help determine those attributes of a graph drawing that suggest that it is human-like or machine-like.

\section*{Acknowledgements}
We are grateful to all the experimental participants, to
Drew Sheets who assisted with creating the graphs in yEd, and to John Hamer who implemented the online experimental system. Ethical approval was given by the University of Arizona Institutional Review Board (ref: 1712113015). This work is supported by NSF grants CCF-1740858, CCF-1712119, DMS-1839274, and FWF grant P 31119.

\bibliographystyle{splncs04}


\newpage
\appendix
\section{Example graph in all eight versions}\label{app:Example}
Graph number 4 ($G_4$) in the experiment shown below in all eight versions. All the experimental stimuli can be found in the supplementary material (visit \url{http://www.dcs.gla.ac.uk/~hcp/GD2020}).

\begin{table}[h]
\caption{Graph number 4 in all eight versions.}
\centering
\begin{tabular}{|c|c|c|c|}
\hline
\includegraphics[scale=0.06]{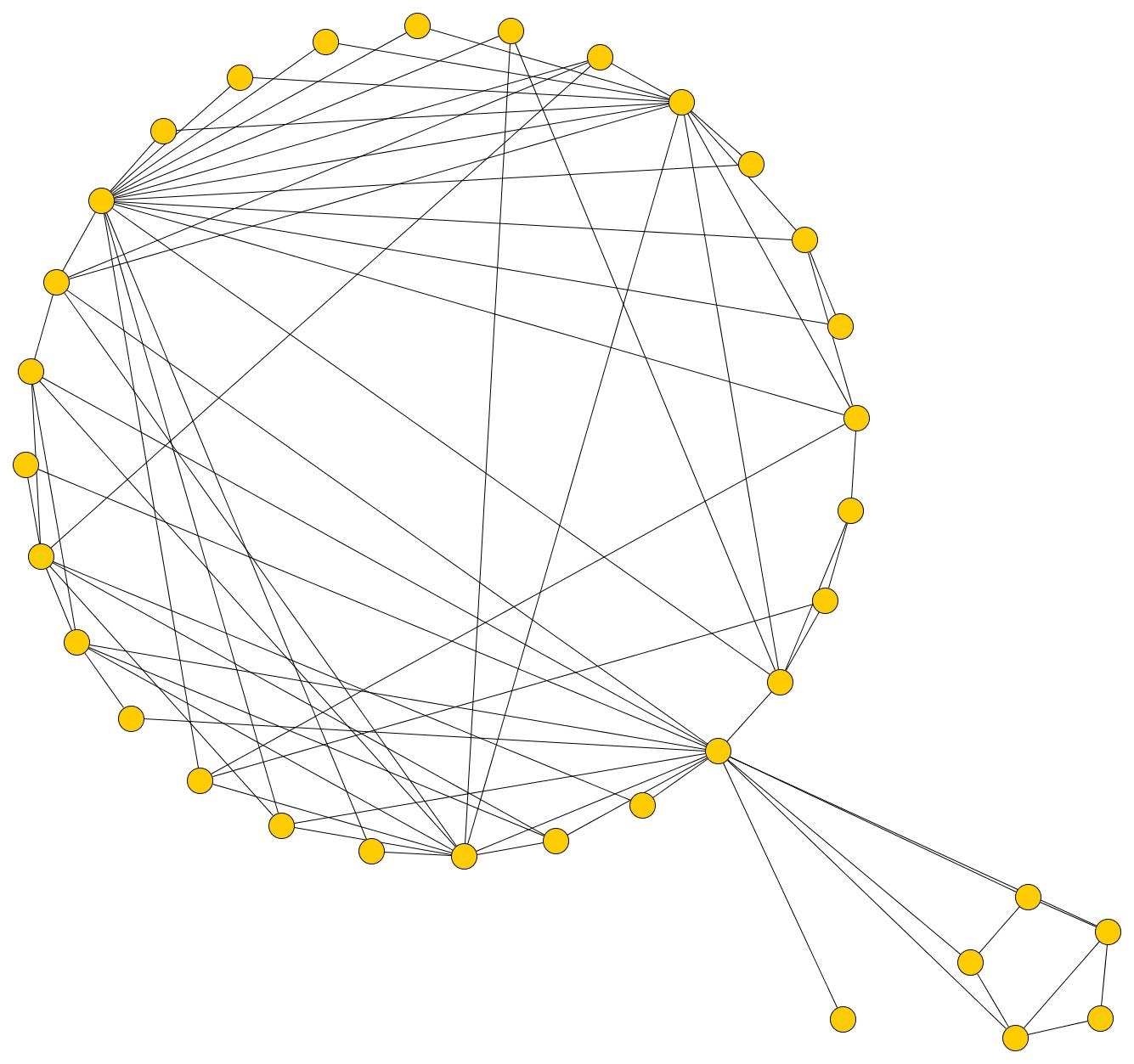} &
\includegraphics[scale=0.06]{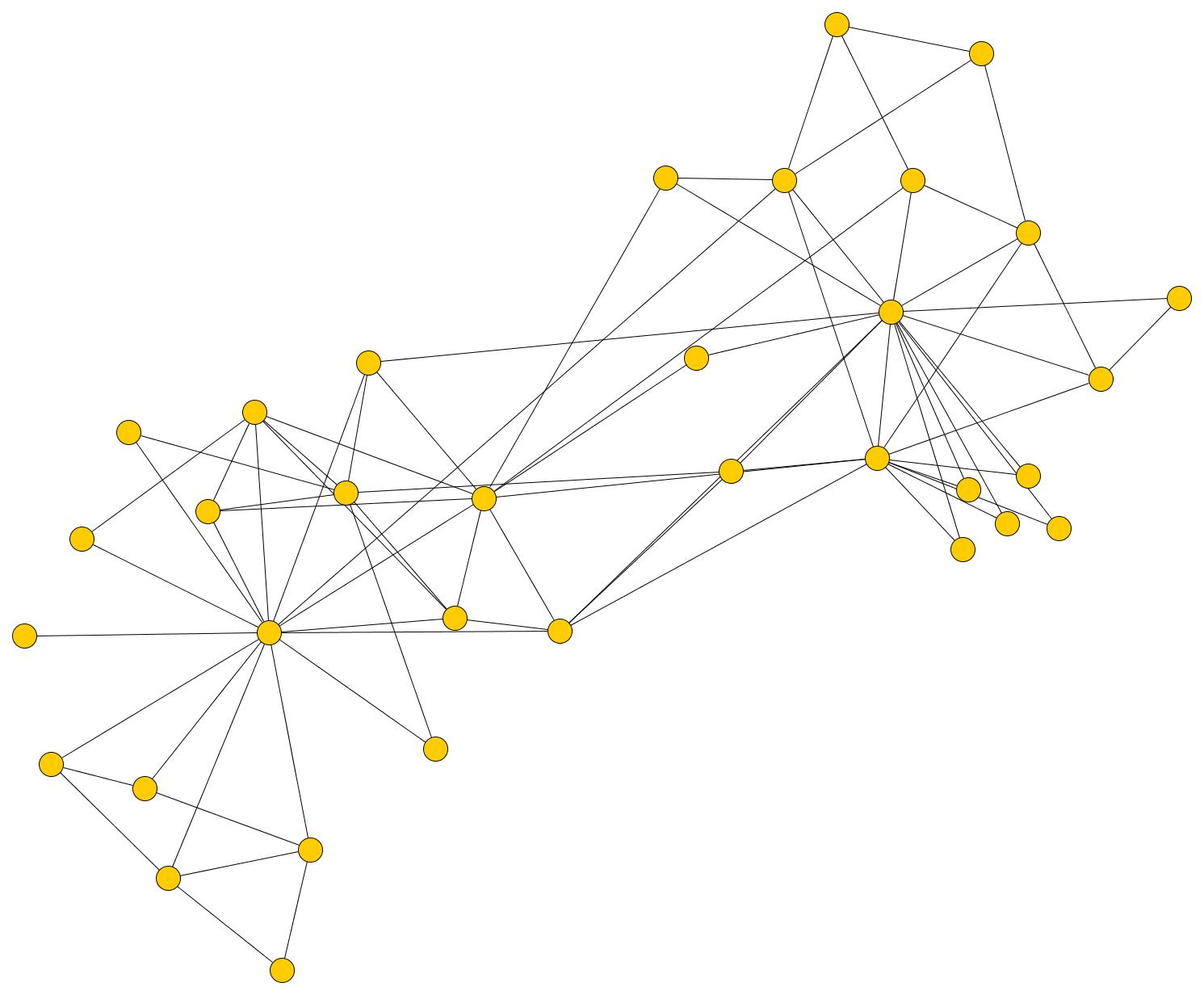} &
\includegraphics[scale=0.06]{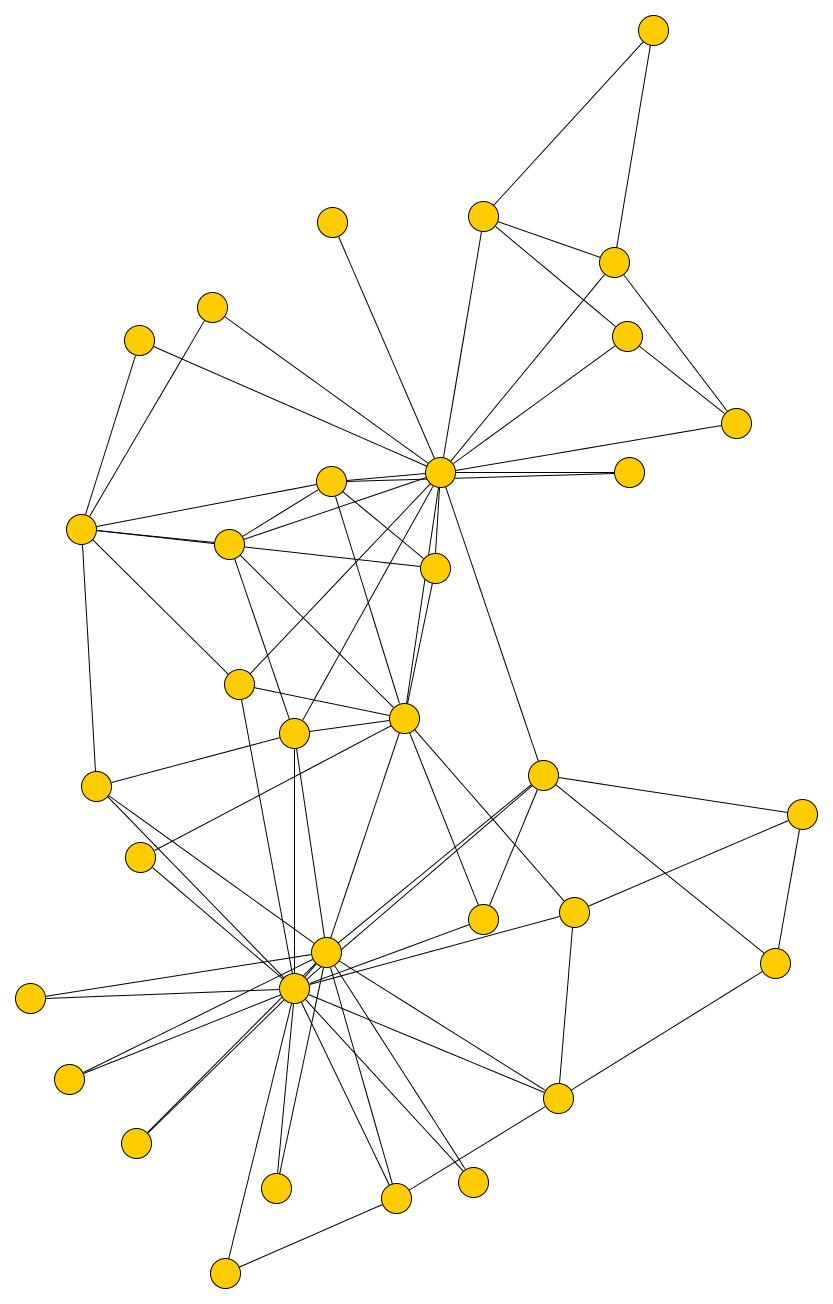} &
\includegraphics[scale=0.06]{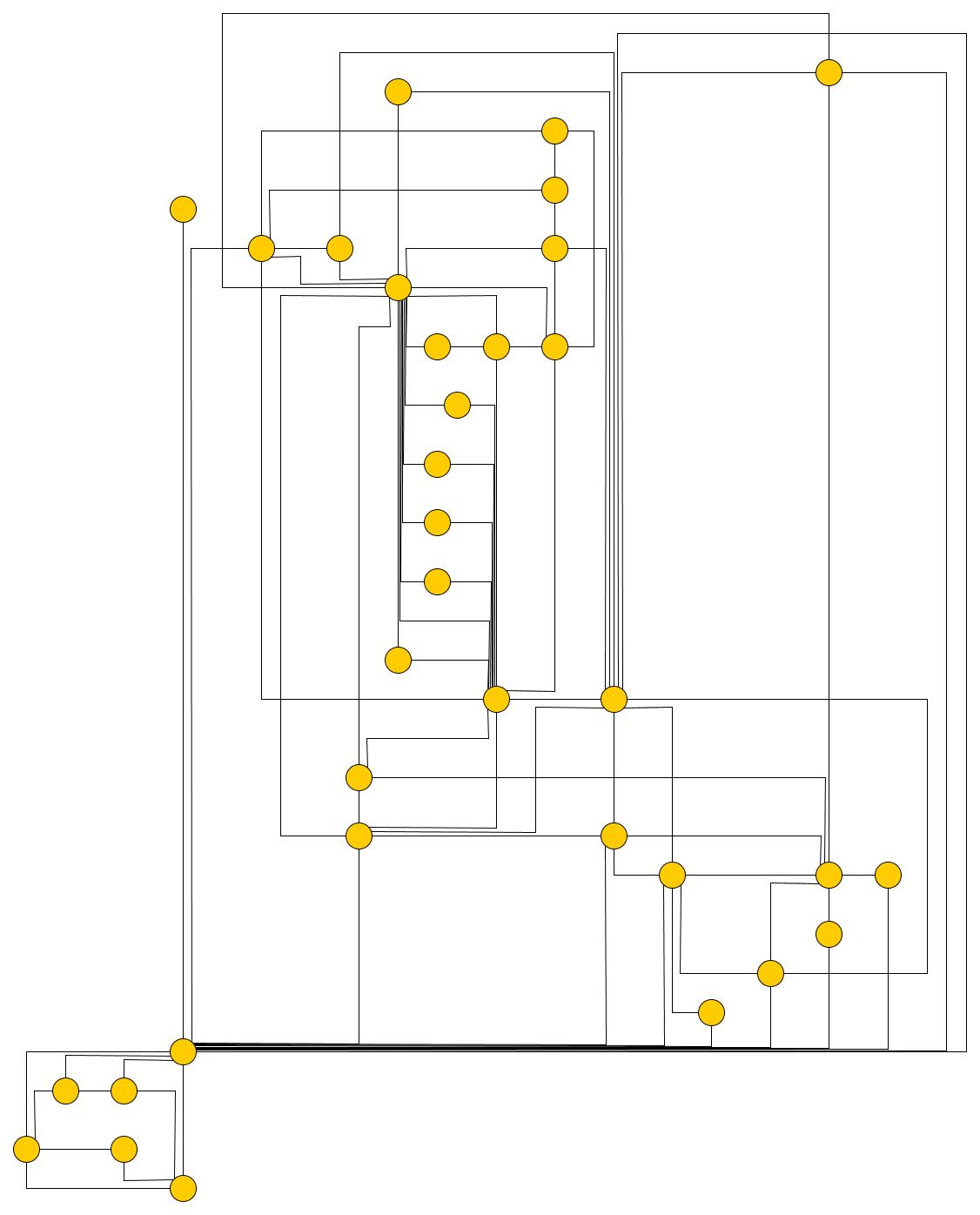} \\
\hline
Circular & Force Directed & Multi-dimensional Scaling & Orthogonal \\
\hline
\includegraphics[scale=0.06]{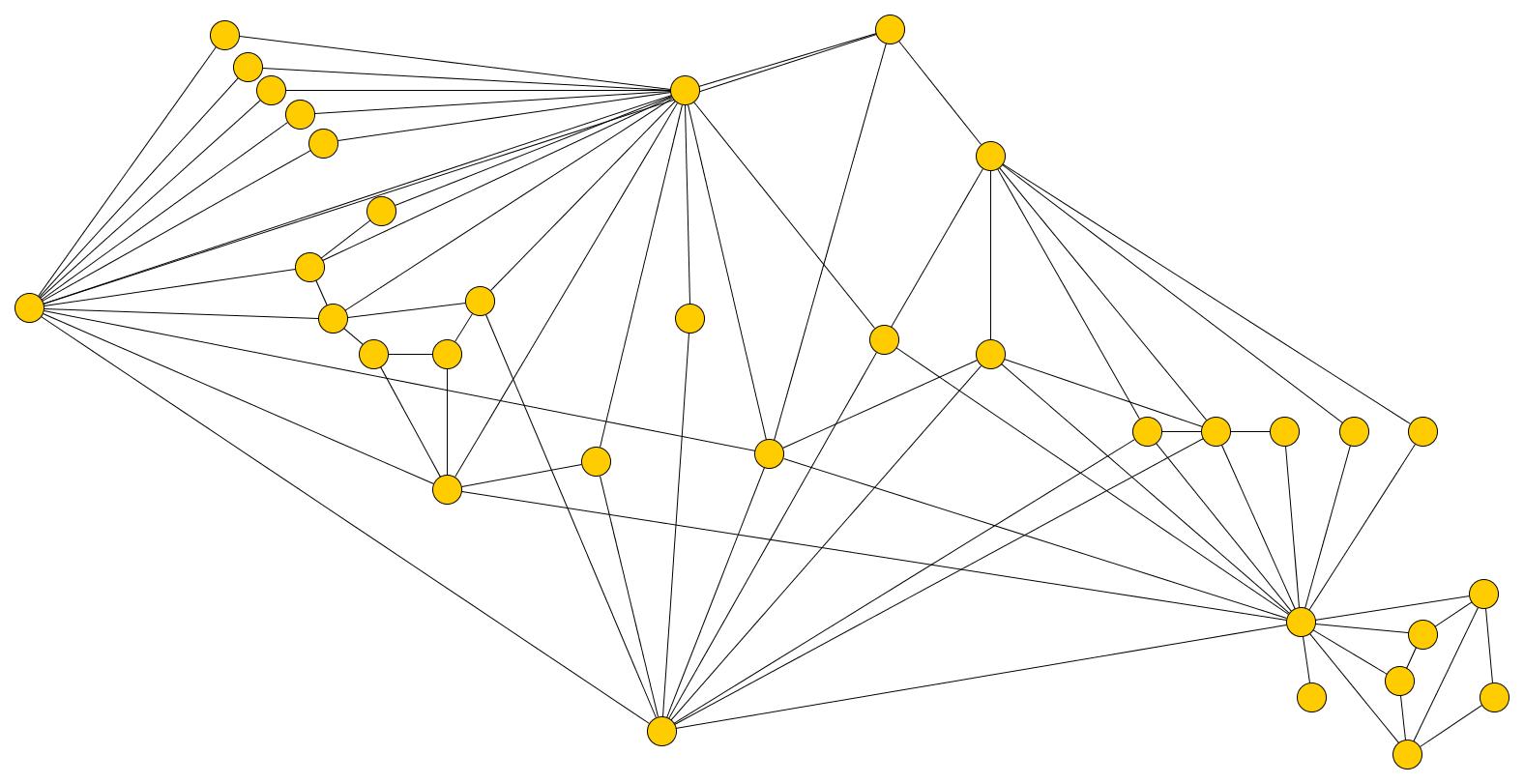} &
\includegraphics[scale=0.06]{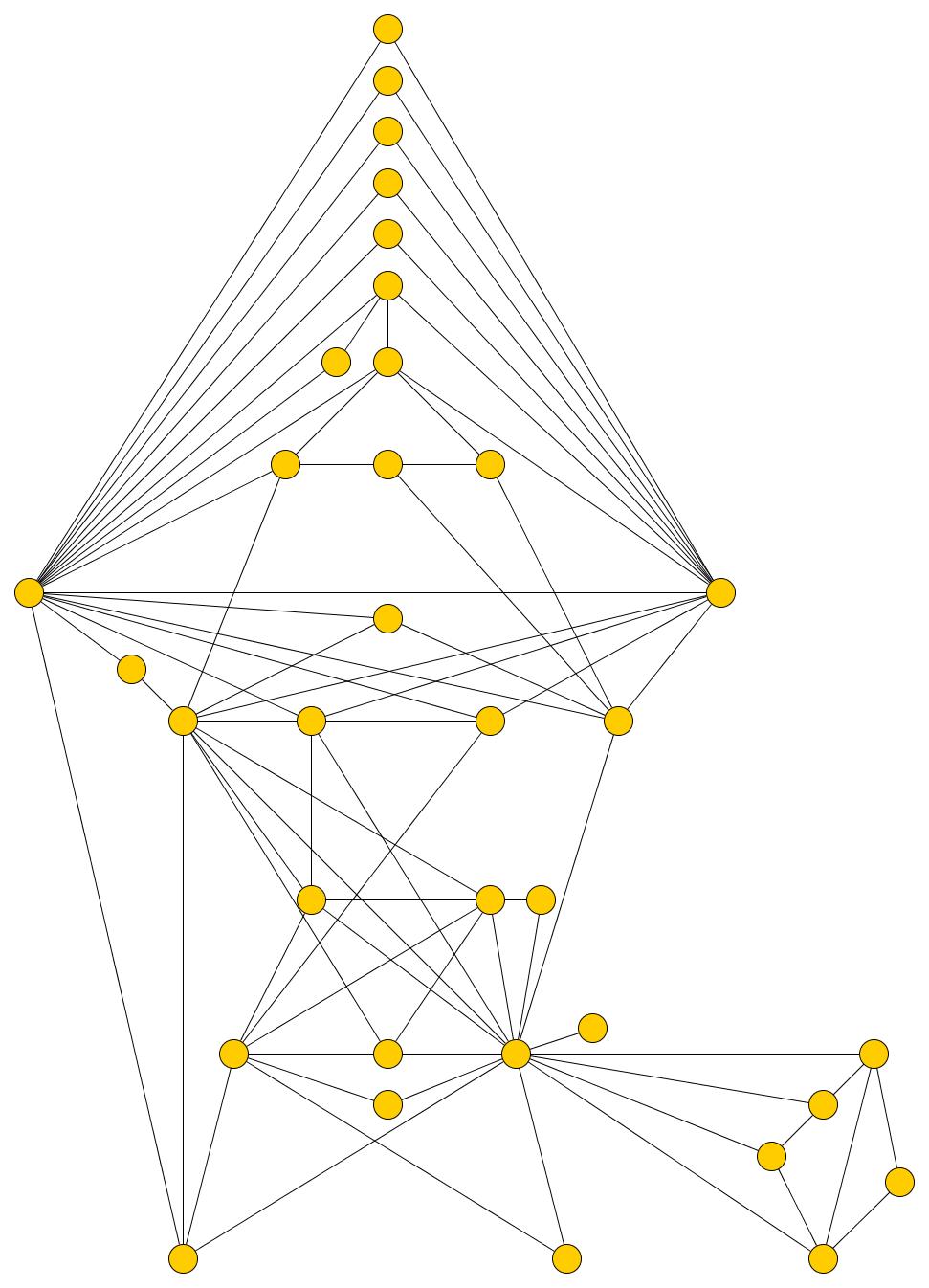} &
\includegraphics[scale=0.06]{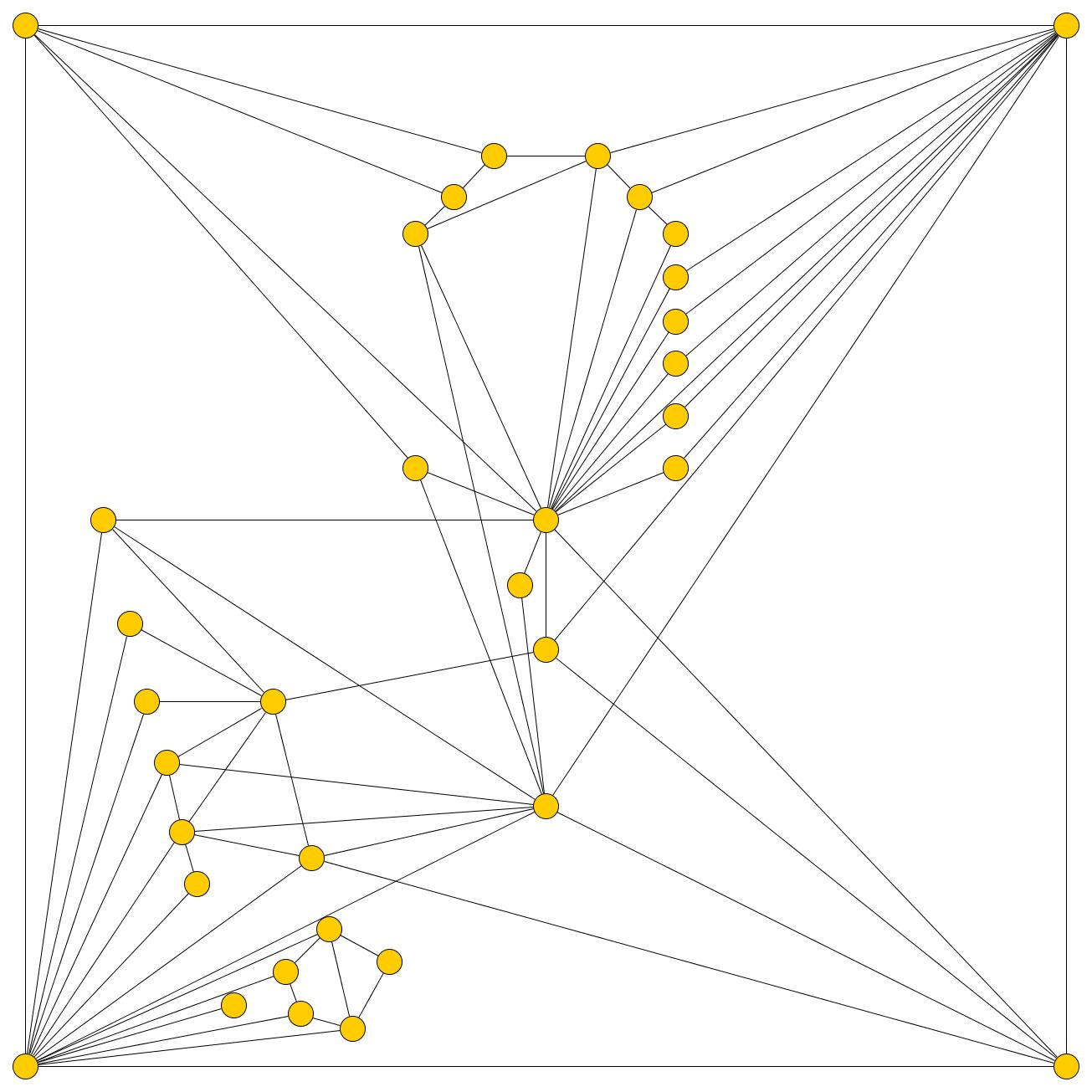} &
\includegraphics[scale=0.06]{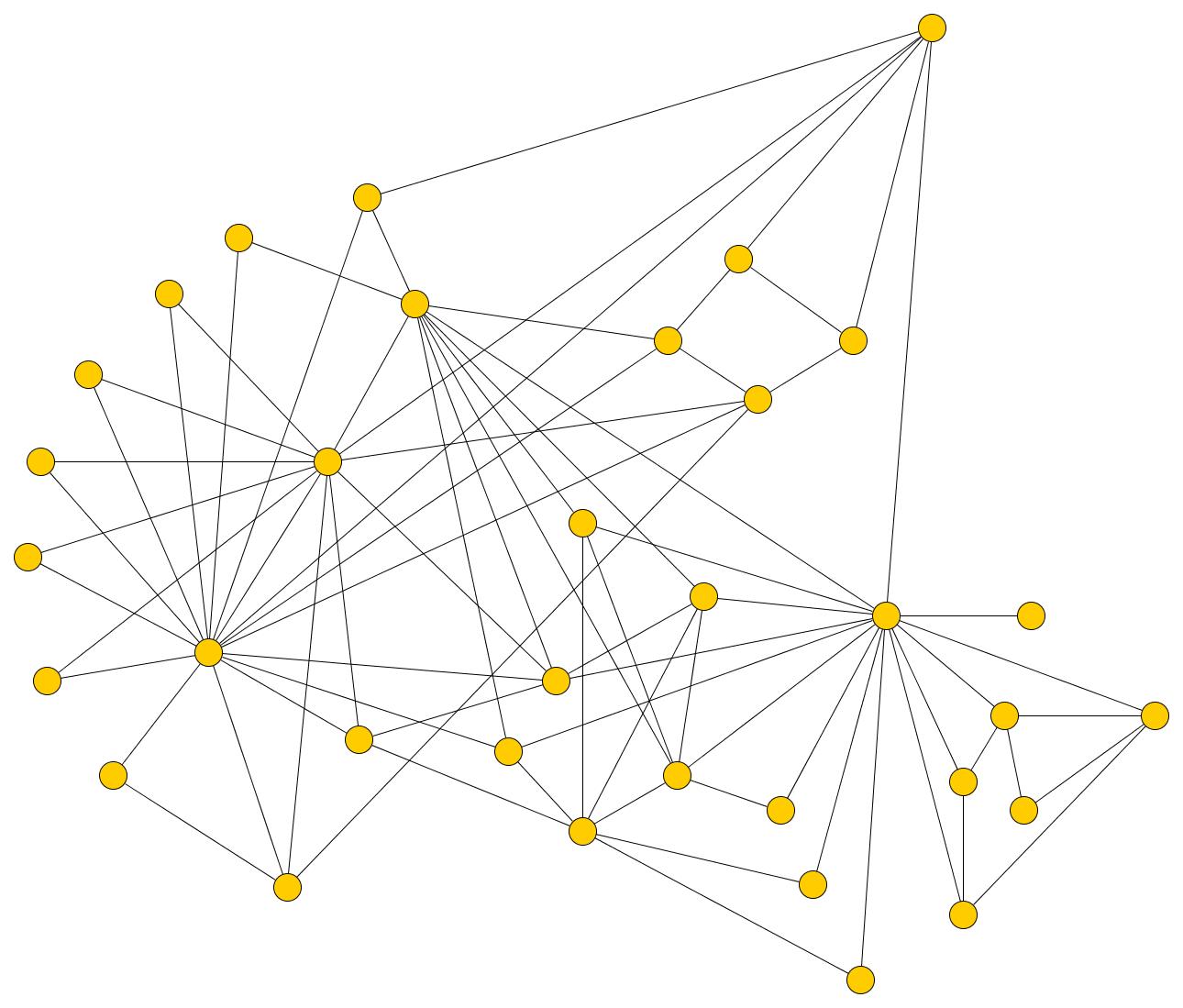} \\
\hline
Human Drawer 1 & Human Drawer 2 & Human Drawer 3 & Human Drawer 4\\
\hline
\end{tabular}
\end{table}

\newpage
\section{Demographics}
\label{app:Demographics}

\begin{center}
\begin{figure}
    \centering
  \includegraphics[width=0.7\linewidth]{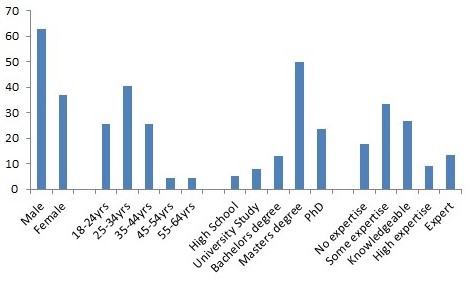}
  \caption{Distribution of demographic information of our participants in the experiment.}
  \end{figure}
\end{center}
 
 \section{Time Taken for Human Drawing}
 \label{app:time}
 
 The drawers were asked to note the length of time taken to draw each graph; one drawer, $D_3$, did not note the length of time, but said that drawing all nine graphs took over $24$ hours.
 
\begin{table}
\caption{Length of time taken to draw the graphs, in minutes}
\centering
\setlength\tabcolsep{6pt}
\begin{tabular}{lrrrrrrrrr} \toprule
		& $G_1$	& $G_2$	& $G_3$	& $G_4$	& $G_5$	& $G_6$	& $G_7$	& $G_8$	& $G_9$	\\ \midrule
$D_1$	& 42	& 9		& 27	& 15	& 12	& 5		& 17	& 9		& 12	\\ 
$D_2$	& 74	& 5		& 53	& 37	& 10	& 12	& 23	& 20	& 33	\\ 
$D_4$	& 36	& 50	& 40	& 19	& 18	& 4		& 15	& 12	& 34	\\ \midrule 
\textit{mean}	& 50.7	& 21.3	& 40.0	& 23.7	& 13.3	& 7.0	& 18.3	& 13.7	& 26.3	\\ \bottomrule
\end{tabular}
 \end{table}

\end{document}